\documentclass{article}

\usepackage{arxiv}

\usepackage[utf8]{inputenc} 
\usepackage[T1]{fontenc}    
\usepackage{hyperref}       
\usepackage{url}            
\usepackage{booktabs}       
\usepackage{amsfonts}       
\usepackage{nicefrac}       
\usepackage{microtype}      
\usepackage{amsmath,amsfonts}
\usepackage{algorithmic}
\usepackage{algorithm}
\usepackage{array}
\usepackage[caption=false,font=normalsize,labelfont=sf,textfont=sf]{subfig}
\usepackage{textcomp}
\usepackage{stfloats}
\usepackage{url}
\usepackage{verbatim}
\usepackage{cite}
\usepackage{multirow}
\usepackage{subfig}	
\usepackage{graphicx}
\usepackage{doi}

\title{Advanced Routing Algorithms for General Purpose Photonic Processors}

\author{ \href{}{\includegraphics[scale=0.06]{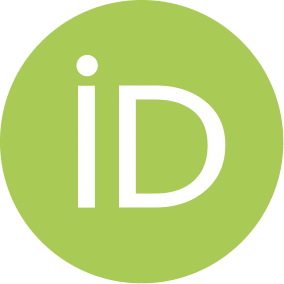}\hspace{1mm}Tushar Gaur}\thanks{} \\
	Department of Electrical Communication Engineering\\
	Indian Institute of Science,\\
	Bangalore, Karnataka,India - 560012 \\
	\texttt{tushargaur@iisc.ac.in} \\
	\And
	\href{}{\includegraphics[scale=0.06]{orcid.png}\hspace{1mm}Gopalkrishna Hegde} \\
	Department of Aerospace Engineering\\
	Indian Institute of Science,\\
	Bangalore, Karnataka,India - 560012 \\
	\texttt{gopalkrishna@iisc.ac.in} \\
	\And
	\href{}{\includegraphics[scale=0.06]{orcid.png}
	\hspace{1mm}Talabatulla Srinivas}\thanks{} \\
	Department of Electrical Communication Engineering\\
	Indian Institute of Science,\\
	Bangalore, Karnataka,India - 560012 \\
	\texttt{tsrinu@iisc.ac.in} \\
}

\renewcommand{\shorttitle}{\textit{arXiv} Template}

\hypersetup{
pdftitle={A template for the arxiv style},
pdfsubject={q-bio.NC, q-bio.QM},
pdfauthor={David S.~Hippocampus, Elias D.~Striatum},
pdfkeywords={First keyword, Second keyword, More},
}

\begin{document}
\maketitle

\begin{abstract}
	Cost-effective and programmable photonic-driven solutions like electronic counterparts (FPGAs) can be implemented using waveguide mesh architectures along with tunable couplers for routing to implement general-purpose photonic processors. These processors/ networks are represented using undirected weighted graphs, where weights are included to implement constraints in the routing. Faster automated routing and cycle finding algorithms are crucial for dynamic path allocations in live networks to implement various functionalities using these processors. We propose path and cycle finding algorithms based on bidirectional and depth-first search techniques, considering various performance metrics for each device to optimize the path according to the required metric. Multiple cases of path distribution and implementation of cycles of various sizes have been demonstrated. Various methods to eliminate the non-functioning or malfunctioning units are proposed. The broad applicability of the proposed path-finding algorithm has been demonstrated using the same algorithm to create a list of all the possible input-output combinations in a 4$\times$4 photonic switching network. A comparison of available search algorithms in terms of execution time and complexity has been described.
\end{abstract}


\section{Introduction}
A communication network integrates multiple functionalities such as modulation, filtering, frequency discrimination, integration differentiation etc., to transmit data. To upscale the data rate, signal processing speed and simultaneously lowering power consumption, onchip multiprocessing cores allowing parallel operations set a new trend in the commercial electronic market \cite{ref1}, \cite{ref2}. However with the advances made in the field of communication realization of scalable on chip electronic communication network faced a critical challenge in terms of bandwidth capacities and electromagnetic interference (EMI), which was solved by switching to the optical communication \cite{ref3}, \cite{ref4}. Initially, bulky optical fiber based systems were used to scale up the network performance, which were replaced by scalable on-chip photonic integrated circuits (PICs) in the last decade \cite{ref5} - \cite{ref8}. PICs offer large bandwidth, low EMI, low losses, small footprints and high power efficiency, thus solving the problems associated with electronics and fiber-based networks. However, today most of the PICs available are application-specific (ASPICs). Thus, to implement multiple functionalities, different ASPICs are connected through optical interconnects leading to high interconnection losses and a larger footprint \cite{ref9}. ASPICs are fabricated using a cost-sharing mechanism, where multiple projects from different users with different designs are fabricated on the same wafer, leading to a high development period \cite{ref10}. Another drawback associated with ASPICs is the non-recon

urability of these devices leading to a new fabrication each time the functionality changes even a little. In the era of IoT (internet of things), to meet modern communication demands, PICs need to follow the footprints of their electronic counterparts. Recently, a reconfigurable general purpose photonic processor (GPPP) was demonstrated as programmable photonic integrated circuits (PPICs) for implementing multiple functions on the same chip\cite{ref11}-\cite{ref13}. 

PPICs can be broadly categorized into two categories: $(i)$ Feed-forward PPICs used to implement arbitrary matrix operations \cite{ref14}, \cite{ref15} and $(ii)$ Feedback PPICs implemented using tunable couplers acting as routers and waveguide meshes, arranged in a particular topology (rectangular, triangular or hexagonal)\cite{ref12}, \cite{ref16}. Later is used as GPPP as they allow the implementation of densely integrated resonant and non-resonant architectures. Light distribution in these circuits is governed by manipulating the phase of tunable couplers (generally implemented using MZIs due to enhance fabrication tolerance) to reconfigure the circuit for implementing multiple functions parallelly, thus enabling parallel processing. Implementation of a required functionality in these circuits requires the following steps:
\begin{enumerate}
    \item \textbf{Mapping: }Given the properties of coupler and interconnection waveguides, find how many units (couplers) are required to implement a given functionality.
    \item \textbf{Routing: }Performs path searching between inputs and outputs to satisfy the criteria obtained in the mapping.
    \item \textbf{Optimization: }Optimizes the individual device performance in the obtained route to maximize the required output \cite{ref17}. 
\end{enumerate}
This article is only limited to routing part among these steps. To program a function in these circuits, user can manually choose the path between the available set of inputs and outp-\\
\\
uts. However, this method becomes cumbersome in cases where large networks are employed and in processes demanding fast reconfiguration. To resolve this issue, these circuits can be represented using graph networks where a node or a collection of nodes along with edges represent a device in the graph. Thus enabling graph and congestion-solving
algorithms to automate the path search in the circuit for implementing the required functionality. Routing and congestion-solving in an integrated photonic network utilize concepts and strategies similar to electronic networks \cite{ref18} but with different constraints due to the reciprocal nature of photonic devices.

Recently, authors have reported routing algorithms based on modified Dijkstra's shortest path and heuristics approach for feedback PPIC networks \cite{ref19}, \cite{ref20}. However, routing using a modified depth first search (DFS) algorithm has not been explored in these circuits yet. DFS algorithm \cite{ref21} - \cite{ref23} is capable of handling complex graphs with negative cycles, more suitable for decision-making, and faster, especially in an extensive network \cite{ref24}. Also, all path and cycle search algorithms crucial for implementing hash lists and resonant architectures in GPPPs have not been discussed previously. Thus, in this paper, we propose DFS based all path, shortest path and cycle search algorithms for GPPP and arbitrary photonic switching networks. The proposed all path search algorithm enables the user to search all the existing paths between a source-target pair, thus allowing the implementation of hash list. For dynamic path allocation in live circuits, the speed of path search becomes a critical criterion. The proposed shortest path search algorithm's results allow speedy path allocation as the search speed scales up during multipath search. Also, in the case of signal processing, cycle search becomes essential to implement simple and higher-order filters based on resonant architectures, and the proposed cycle search algorithm enables the implementation of resonant architectures in GPPP. This paper also discusses the faster BFS based modified bidirectional search algorithm, along with its limitation. The broad applicability of the proposed algorithm in an arbitrary PIC network has been reported by applying the proposed algorithm in a N\texttimes N photonic switching network. A comparison with existing algorithms in terms of complexity, longest path searched, and execution time for various networks under different situations has been discussed.   

\begin{figure*}[!t]
\centering
\subfloat[]{\includegraphics[width=2in]{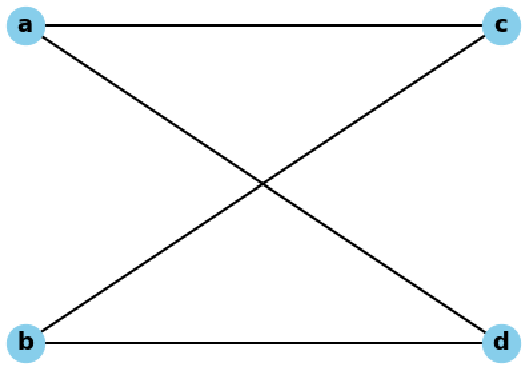}%
\label{fig1a}}
\hfil
\subfloat[]{\includegraphics[width=2in]{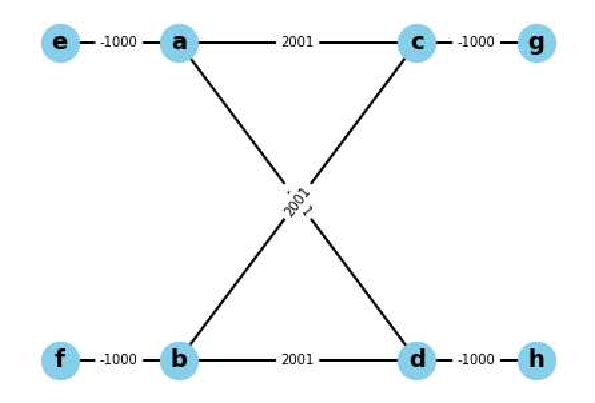}%
\label{fig1b}}
\caption{(a) An MZI represented using a Bipartite graph and (b) Dummy nodes added to bipartite graph at both the inputs and outputs with corresponding weights.}
\label{fig1}
\end{figure*}

\section{Graph Representation}

\subsection{Representation of a Single MZI}
Directional couplers/ MZIs acting as the routers in a general purpose photonic processors (GPPPs) form the fundamental unit of the circuit and can be represented using a directed or undirected bipartite graph as shown in Figure \ref{fig1a}. These MZIs can stay either in a cross or bar, or tunable state and conventional graph search algorithms can be employed to find the paths between the nodes of an individual MZI or a connected network of MZIs. However, searched path will consist of both physical and nonphysical paths. For example, two paths obtained between node 'a' and 'c' are: 'a'-'c', which is a physical path and 'a'-'d'-'b'-'c', which is a nonphysical path as back reflected paths are not allowed/considered in optical devices. Also, an MZI, if assigned a particular state (bar or cross) after searching a path, must remain in the same state for another path as well while searching multiple paths in a network. This is called the bar cross condition. Thus, various methods can be employed to eliminate the nonphysical paths and simultaneously maintaining the bar cross condition while designing the network and algorithms \cite{ref19}, \cite{ref20}:
 \begin{enumerate}
    \item The algorithm can be modified such that same device cannot be accessed twice in a path by setting the maximum number of nodes traversed for same device to 2. This method works fine for finding the shortest path but fails to detect the cycles and paths with the desired length (number of MZIs) in the network. 
    \item The network can be implemented using directed graphs, in this case two layered graphs are needed for accessing the network from both the directions, one for each direction. Also, every time a path is found in any direction, it needs to be engaged in the other layer, to avoid contradictory assignment of states of MZIs while searching multiple paths in a network. It increases the space and time complexity of the path searching algorithms.
    \item Another method is to include the dummy nodes with large negative weights at both the inputs and outputs as shown in Figure \ref{fig1b}. In this case if any nonphysical path is encountered the weight will grow large, and for a physical path the total weight for each traversed unit will be 1. It doubles the space complexity of the graph as compared to simple bipartite graph but increases the accuracy of path search algorithm. 
\end{enumerate}
Penalties such as delay, insertion loss and power consumption can be added as edge weights in each unit while applying any of the above mentioned solutions, and algorithms can be modified to optimize any of these penalties either one at a time or simultaneously.

\begin{figure*}[ht]
\centering
\subfloat[]{\includegraphics[width=2.2in]{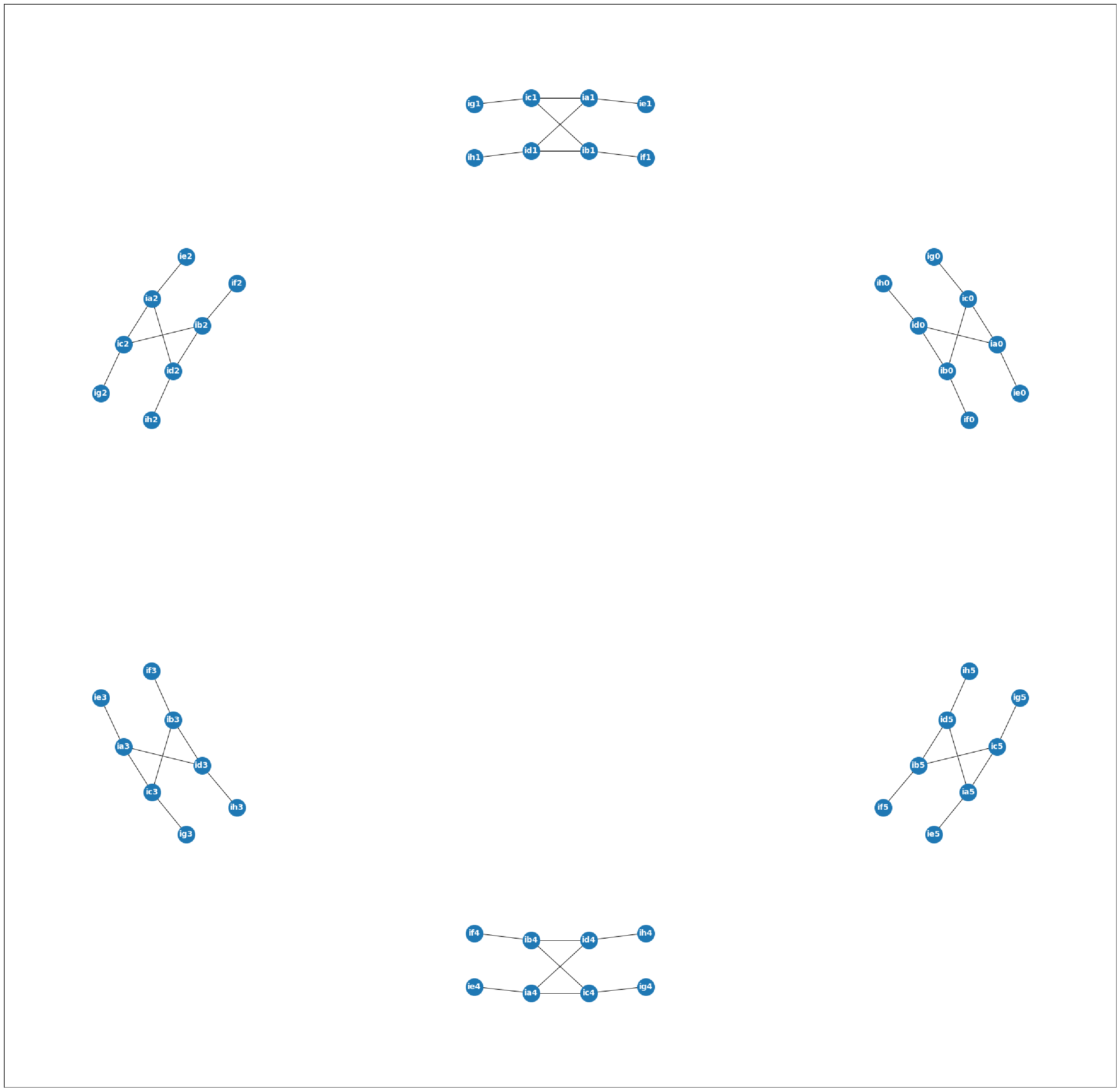}%
\label{fig2a}}
\hfil
\subfloat[]{\includegraphics[width=2.2in]{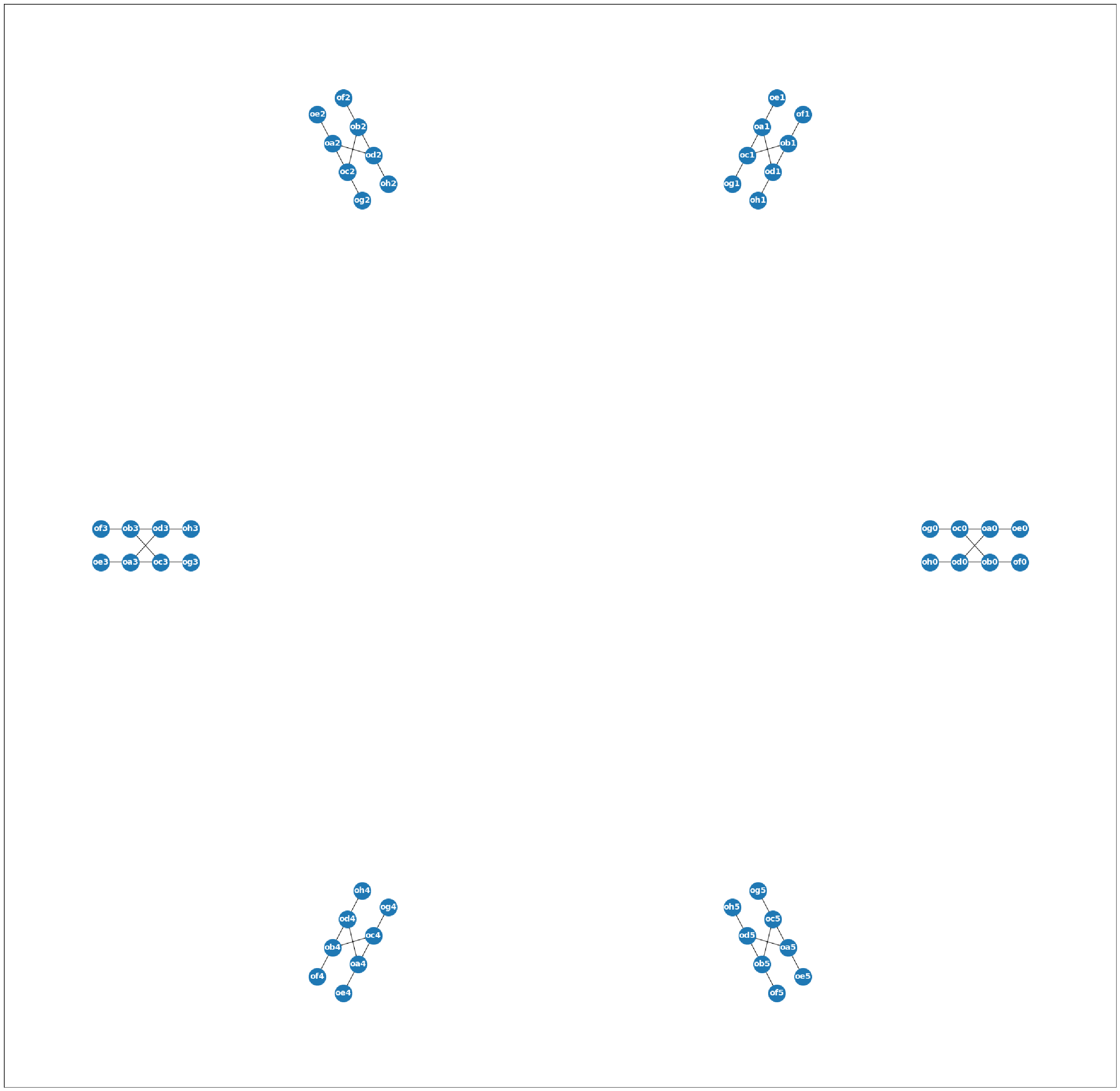}%
\label{fig2b}}
\hfil
\subfloat[]{\includegraphics[width=2.2in]{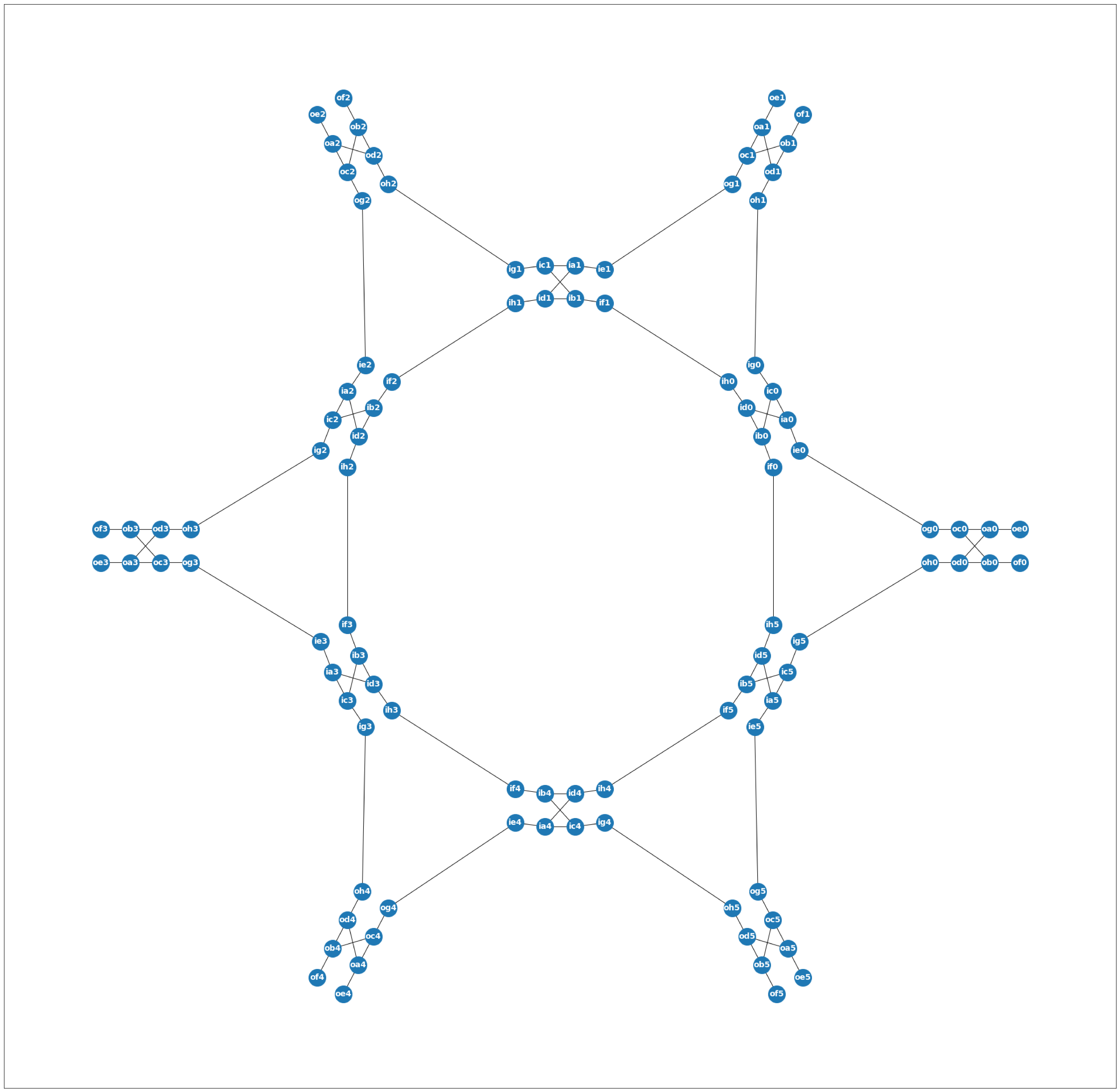}%
\label{fig2c}}
\caption{Fundamental units arranged in (a) inner circle and (b) outer circle, and (c) combination of the two circles forming a unit cell.}
\label{fig2}
\end{figure*}

\begin{figure*}[!t]
\centering
\subfloat[]{\includegraphics[width=2.2in]{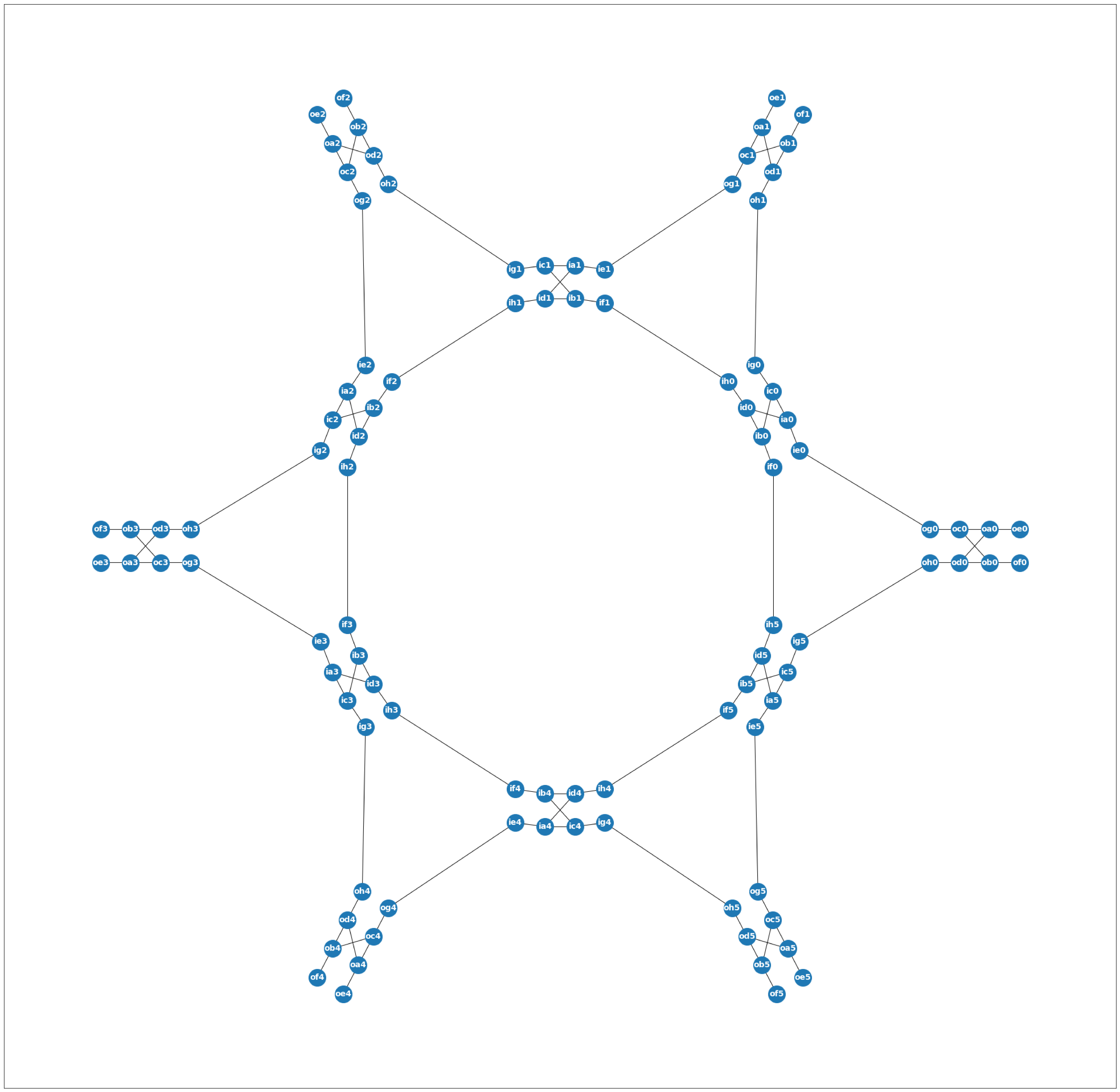}%
\label{fig3a}}
\hfil
\subfloat[]{\includegraphics[width=2.2in]{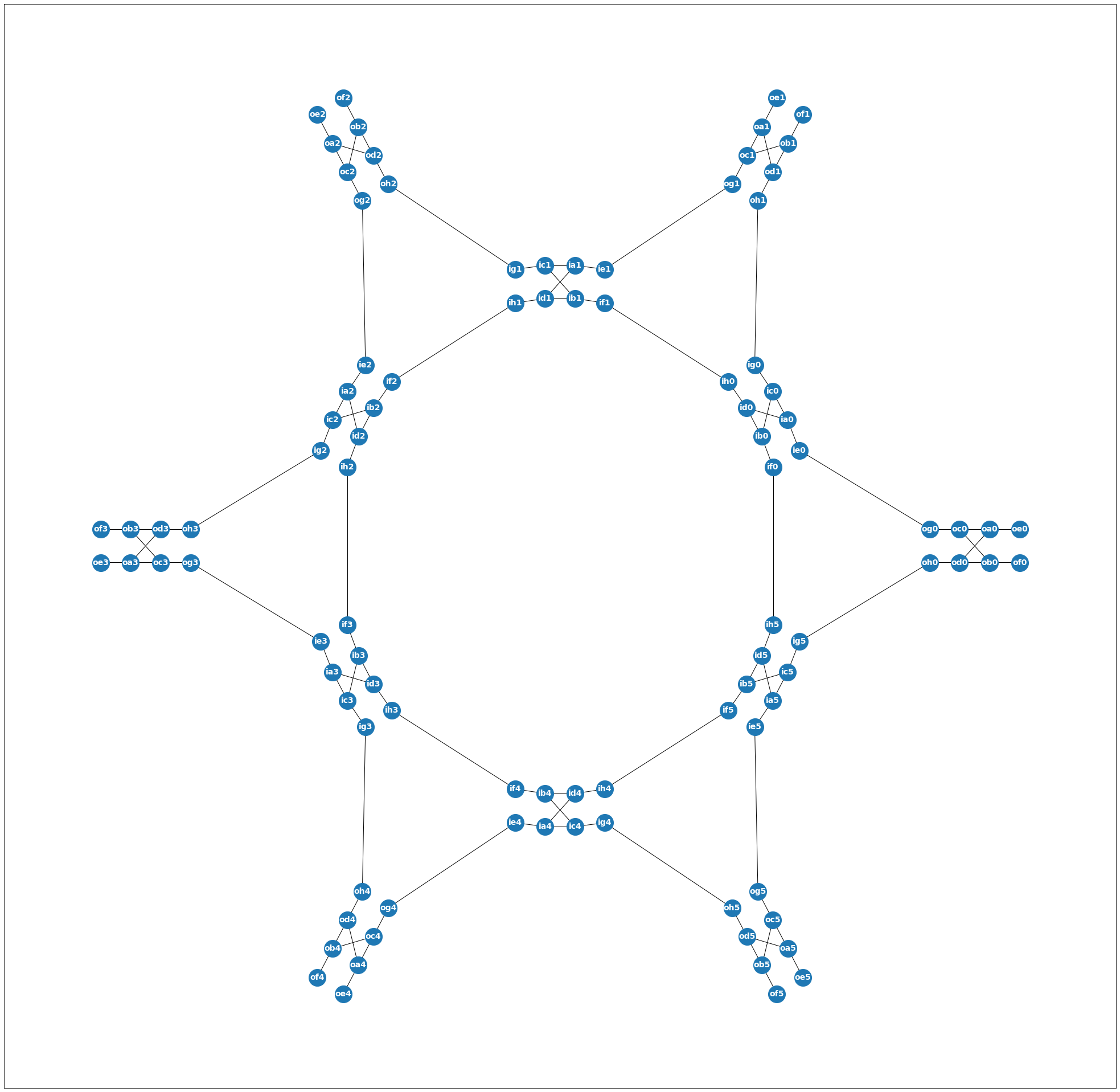}%
\label{fig3b}}
\hfil
\subfloat[]{\includegraphics[width=2.2in]{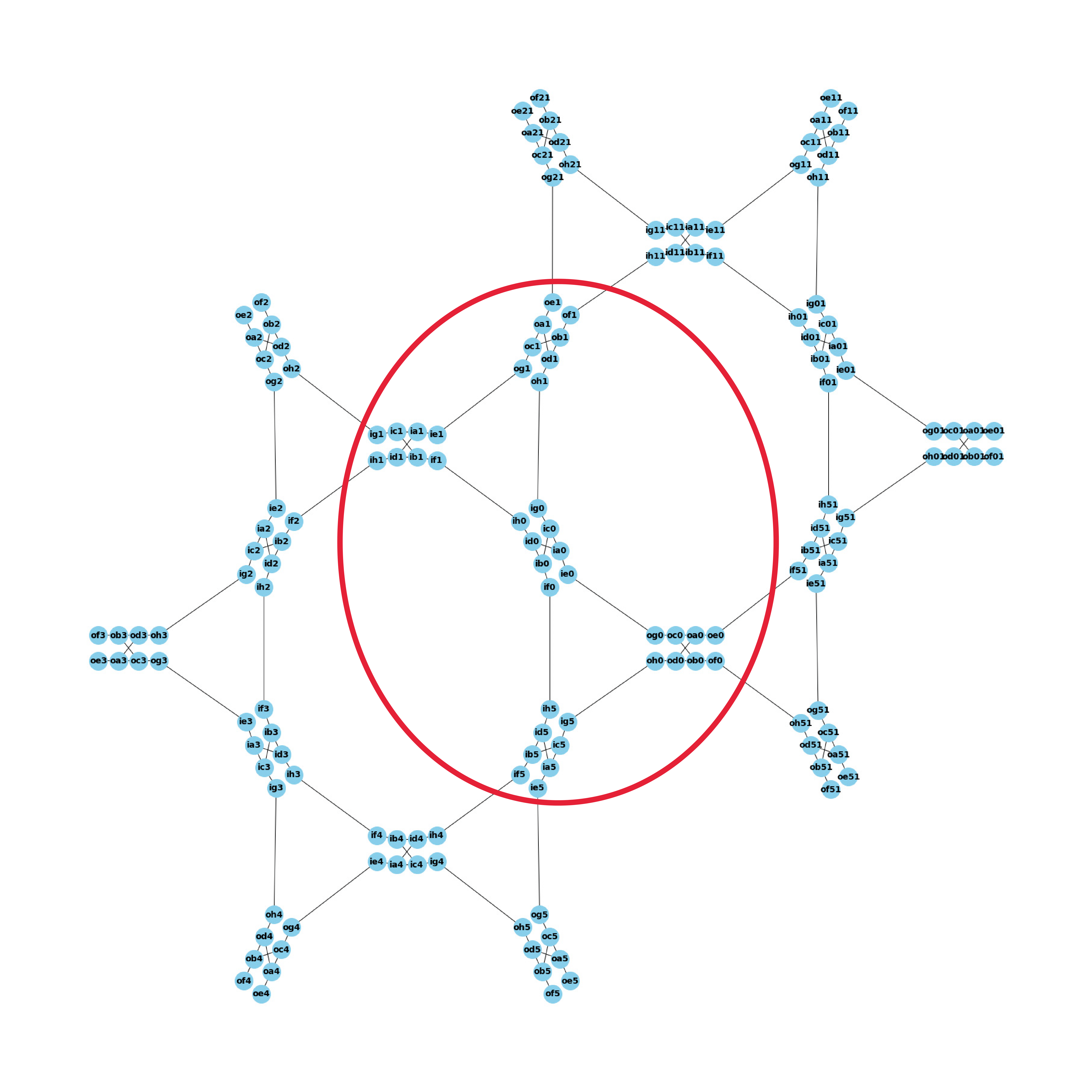}%
\label{fig3c}}
\caption{Combination of 2 unit cells to form a network while deleting the overlapping nodes from one of the cells (overlapping nodes marked in red).}
\label{fig3}
\end{figure*}

\subsection{Network Design}

Given the advantages of hexagonal topology, the hexagonal arrangement of MZIs is utilized to implement the GPPP architecture. The fundamental unit (MZI) is defined using the dummy nodes as shown in Figure \ref{fig1b}. Here large negative and positive weights are allocated to the edges connecting the dummy nodes and primary nodes, respectively. As a result of this allocation, the total weight for a physical path (eg: 'e'-'a'-'c'-'g', 'e'-'a'-'d'-'h') in the MZI becomes 1, whereas the weight for a nonphysical path (eg: 'e'-'a'-'d'-'b'-'c'-'g', 'e'-'a'-'c'-'b'-'d'-'h') becomes 4003. Based on these allocations, as soon as a node of a nonphysical path is encountered in the search algorithm, the total weight becomes $>=3002$. Thus a threshold weight can be assigned for eliminating nonphysical paths in search algorithms.  MZIs are arranged across an inner and outer circle to form a unit cell, as shown in Figures \ref{fig2a} and \ref{fig2b}, respectively. Nodes of MZIs along the inner and outer circles are named using 'i' and 'o', respectively. A unit cell is formed by combining the two circular arrangements as shown in Figure \ref{fig2c}. To implement a multi-cell network, the number of cells required in the network is entered through an input prompt, based on which cells are distributed around the central cell. The Centre positions of these cells are calculated using inner and outer circle radii. Unit cells formed at their respective positions are added to the network, and overlapping nodes between the cells are found and deleted from one cell while connecting the individual cells, as shown in Figure \ref{fig3}. Nodes are numbered according to the cell number starting from 0 for the central cell. For example, a node at the inner MZI at a position 'r' in the cell 'q' is represented as 'ixrq' where 'x' can take alphabets between 'a' and 'h' (referring to fundamental unit's node). Position of each node is saved in a list corresponding to its cell and updated every time a node is added or deleted in the network. A 5-cell and an 11-cell network with 36 and 70 MZIs, respectively are prepared for implementing path and cycle search algorithms.

\section{Algorithms}
\subsection{Shortest Path Bidirectional Search} The bidirectional search algorithm is a BFS based algorithm utilized to find the shortest path in a graph. In this algorithm, two simultaneous searches are performed step by step, one in the forward direction initiating from the start node and another in the backward direction, initiating from the target node, while adding weights at each node. Two visited dictionaries are maintained for search in both directions, and nodes visited in forward and backward searches are added to these dictionaries. To eliminate the nonphysical paths, a condition is added to the search algorithm, and a new node is added to the dictionary only if the weight at that node is smaller than the threshold weight and is not in the visited list. An intersection search is employed in each search step to find the intersection between the two dictionaries. If an intersection is found, the path between the start and target node is created from the intersecting node using predecessors and successors in the forward and backward visited dictionaries, respectively, and the total weight is calculated alongside the path. However, this algorithm works well with non-cyclic simple graphs but poses a problem with cyclic or complex graphs. 

\subsection{Depth First Search Path Finding}
DFS is a recursive algorithm that finds all the paths between a source and a target node. It can also be employed to search the path of a given weight in the network by simple modification in the return condition. Search starts from the source node and recursively calls the searching function at each neighbouring node. A visited set is maintained, holding the nodes being visited and a weight set is created to hold the weight at the current node, which is updated at each searched node. To eliminate the nonphysical paths threshold weight condition is added to the recursive calling loop such that the node is added to the search only if the total weight at the node is less than the threshold and is not in the visited list. Otherwise, the search returns to the previous node, thus eliminating the nonphysical path simultaneously while searching. If the current and target nodes are the same, then a path with respective weight is returned and added to the list of paths.

\subsection{Shortest and Fixed Length Path DFS Search}\label{sec3.3}
DFS algorithm performs path search with higher accuracy and speed in complex and cyclic graphs. As DFS is a recursive algorithm in order to stop the search, it needs to come out of all the loops that the algorithm had run step by step, returning to the nodes it has gone through previously. For DFS to return the shortest path, the algorithm is modified such that once the first path is searched and stored in the path variable along with its weight, the algorithm will check the path weight at each new node. If the weight exceeds the weight of the path previously searched, it will not search further from that node and return to the previous node, and if the weight is smaller than the stored weight, it will carry the search either until the target node is found or weight exceeds the saved weight. Thus, if a new path with a smaller weight is found, the path is updated with the new path alongside its weight. 

A similar modification is also applied to search a path with a given weight. In this case, the first path is updated only when a path with the required weight is found. In order to return the path, recursive calling is stopped by continuously returning the path until return functions roll back to the first node where the calling started. 

\subsection{Cycle Search Algorithm}
However, BFS and Bellman-Ford algorithms support the cyclic graph, but a negative cycle graph limits them. Thus, a recursive search algorithm with a concept similar DFS is employed to implement the cycle search algorithm. A visited and total weight set is created and updated on successful search at each node. Here a parent node is added, and a recursive search is called at each node starting from the parent node. Search at a node progresses if the threshold weight condition is met and the node is not in the visited list, like DFS. If the current node is in the visited list and is the same as the parent node, then a cycle with the respective weight is added to the cycle's list. Otherwise, the search returns to the previous node, simultaneously eliminating nonphysical cycles. To find the shortest and fixed length cycles return conditions are modified using the concept used in section \ref{sec3.3}

\section{Results}
Considering the search cases that may occur in a GPPP, the proposed algorithms are utilized for searching single source-target shortest path, single source-target all path, multiple source-target paths and cycle search from a given parent node. To assess the performance of the algorithms in terms of execution time, mean time with different source-target and parent nodes for path and cycle search algorithms, respectively, has been calculated. 

\subsection{Single Source-Target Shortest Path} 
\begin{figure*}[!t]
\centering
\subfloat[]{\includegraphics[width=2.5in]{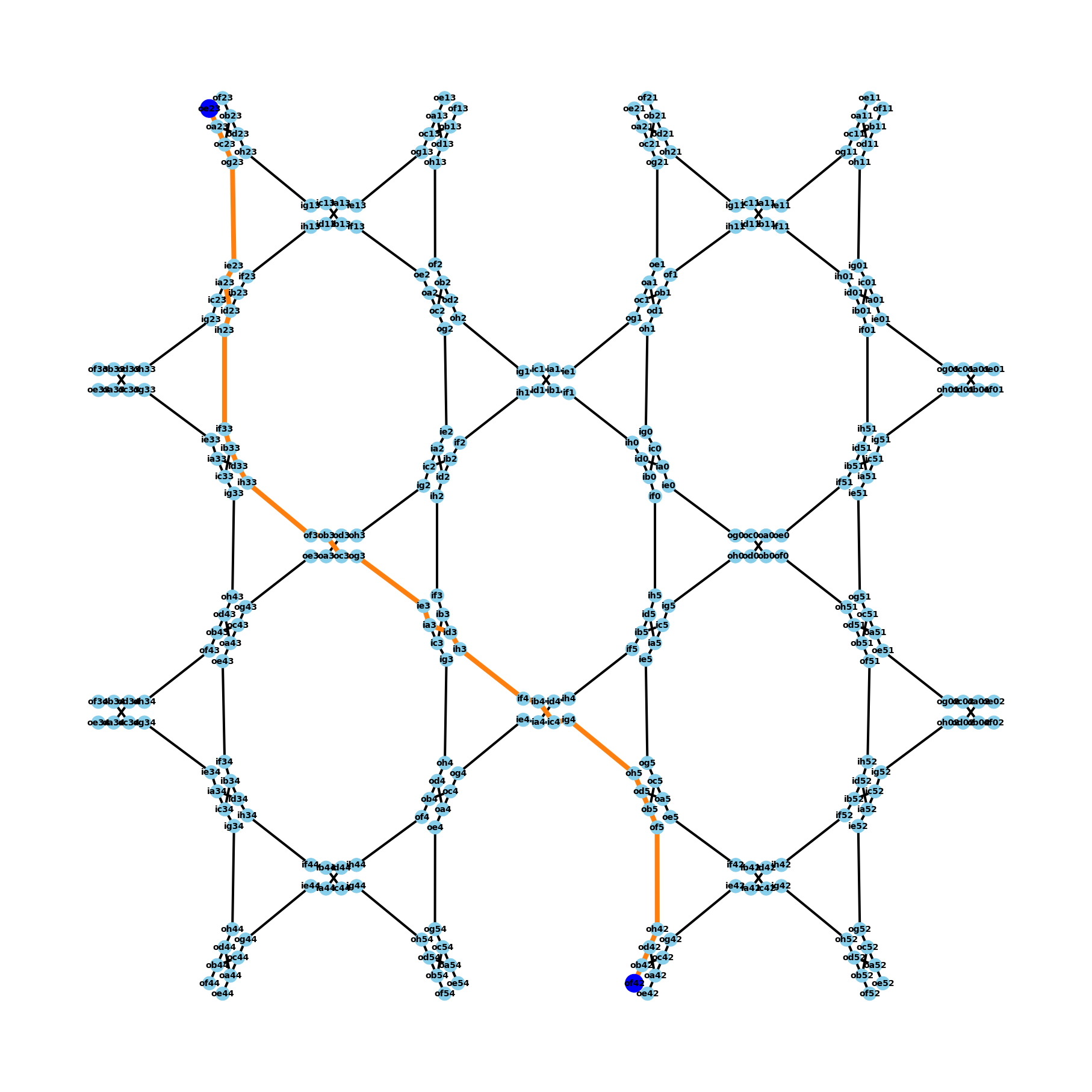}%
\label{fig4a}}
\hfil
\subfloat[]{\includegraphics[width=2.5in]{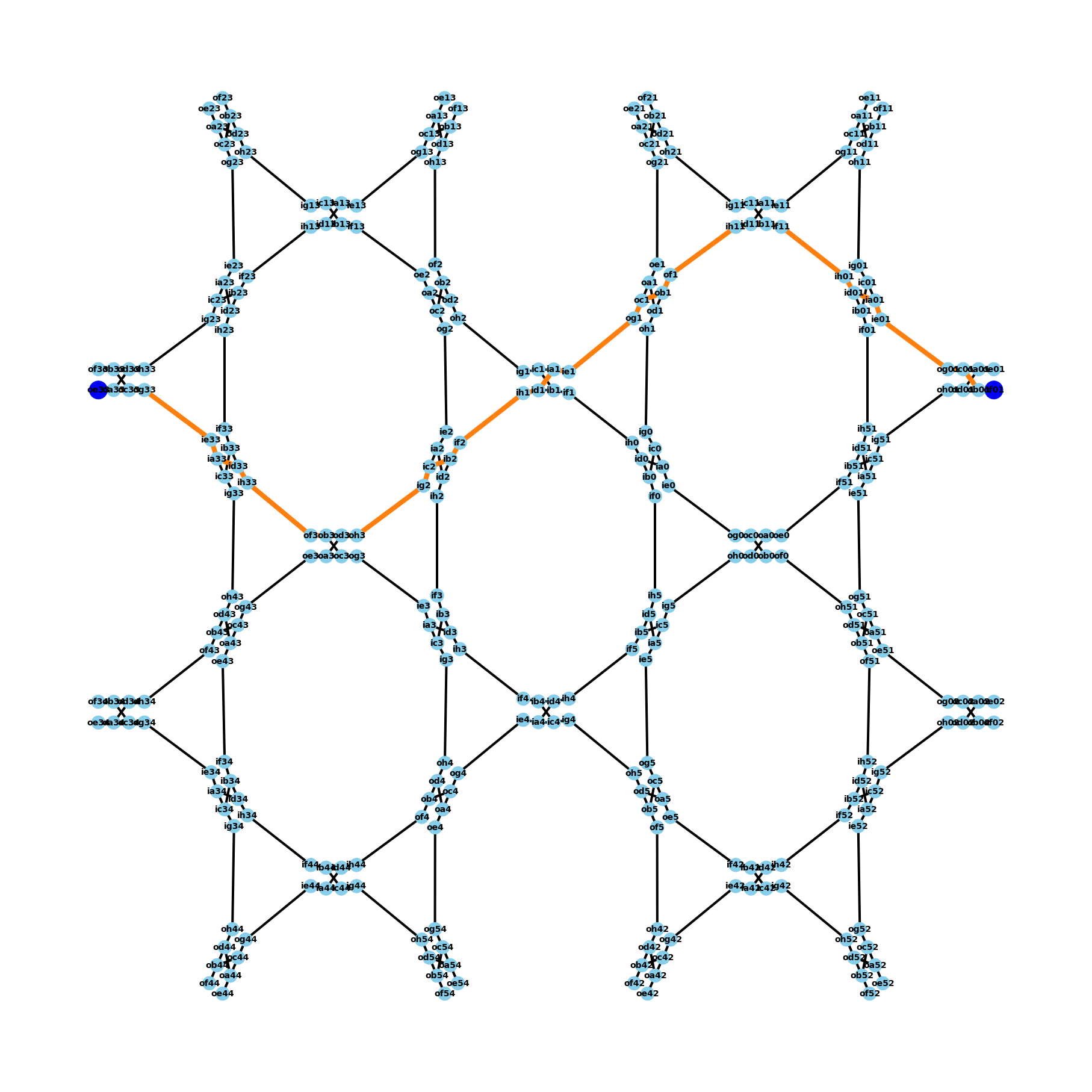}%
\label{fig4b}}
\hfil
\subfloat[]{\includegraphics[width=2.5in]{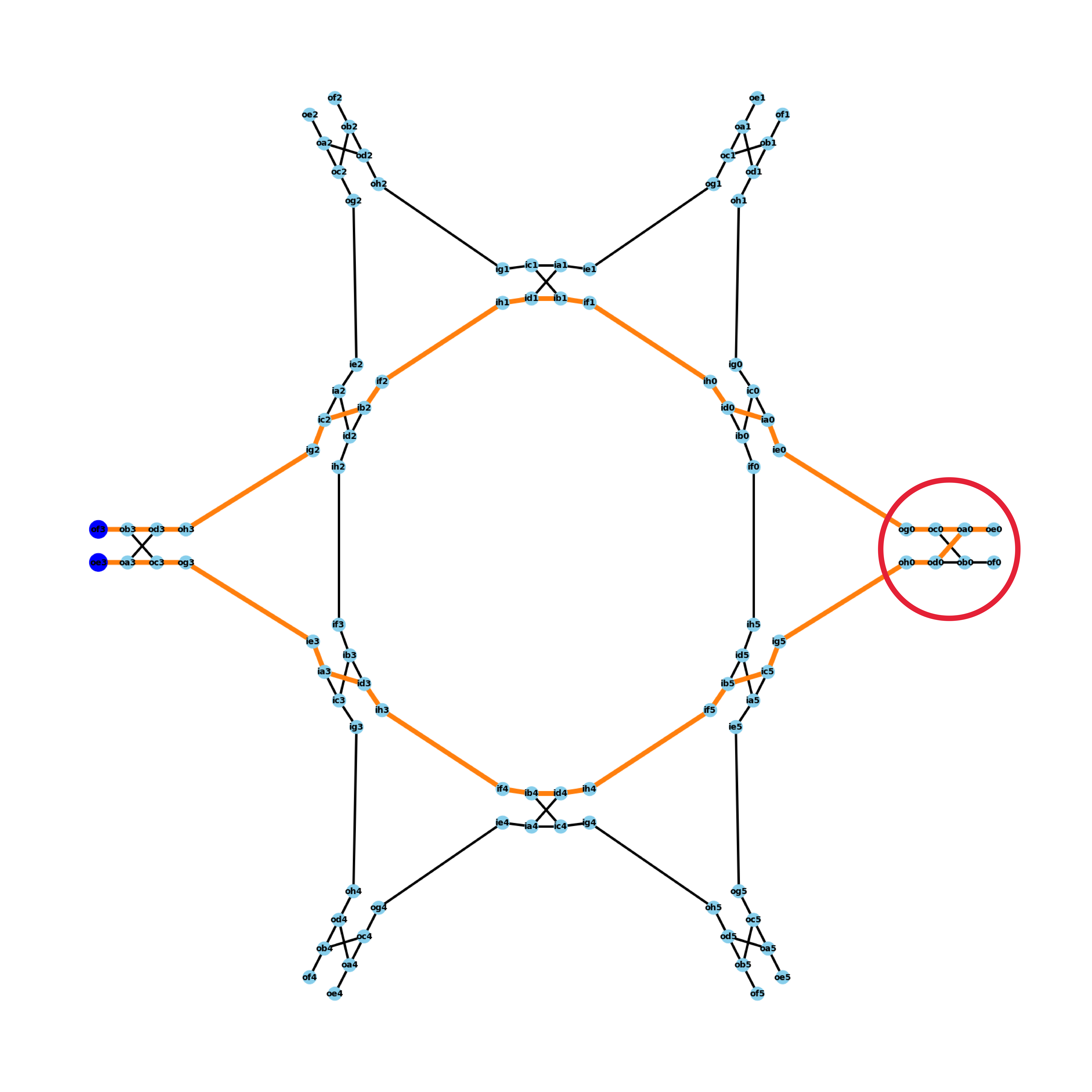}%
\label{fig4c}}
\hfil
\subfloat[]{\includegraphics[width=2.5in]{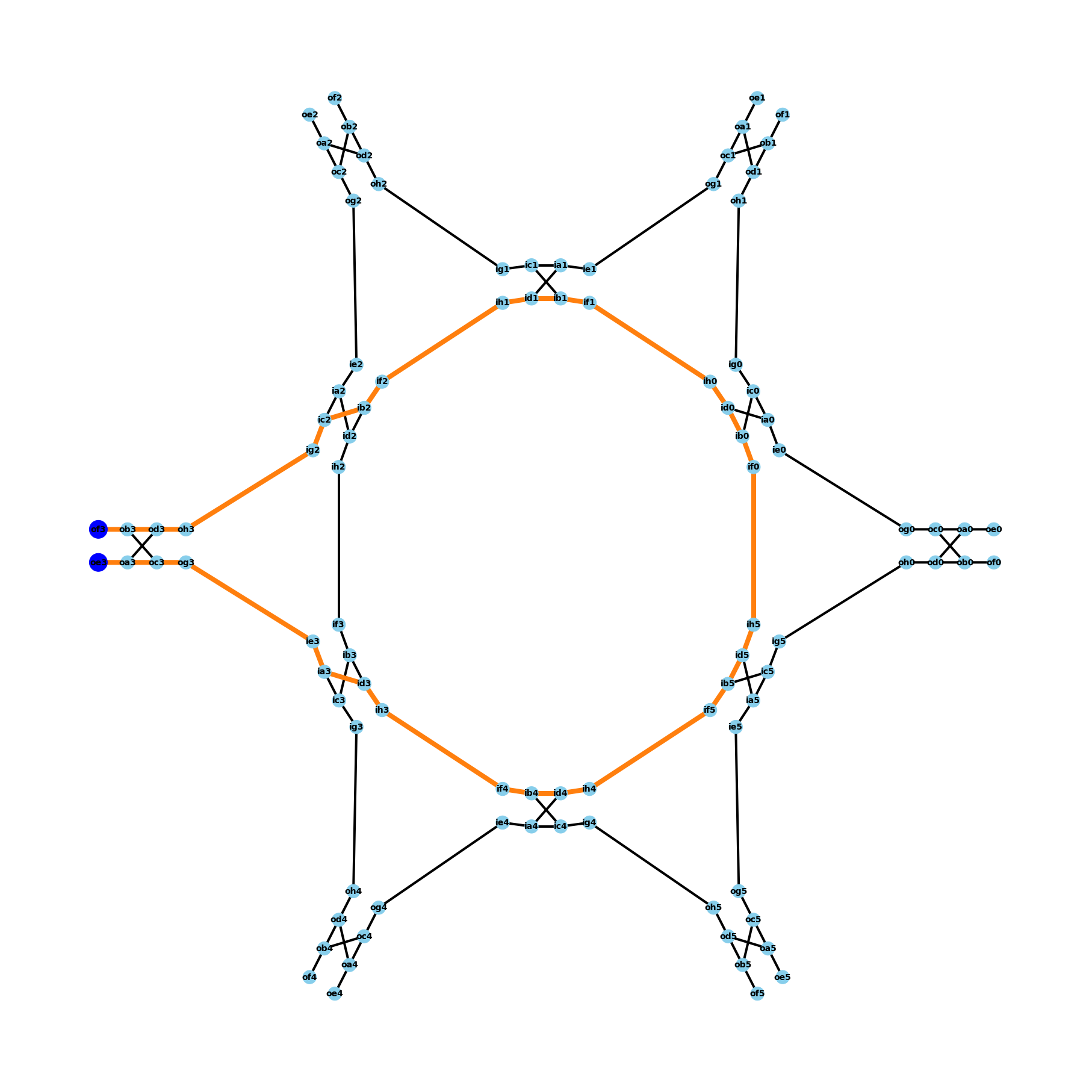}%
\label{fig4d}}
\caption{Single source-target paths (a) 'oe23' to 'of42', (b) 'oe33' to 'of01', (c) 'of33' to 'oe33' case where bidirectional search fails (failure point is marked in red oval), and (d) 'of33' to 'oe33' case where DFS search pass.}
\label{fig4}
\end{figure*}

For searching the shortest path between a pair of source and a target, modified bidirectional search and modified DFS algorithms are employed. A path between nodes 'of33' and 'of01' is found where the total number of travelled units was 9, and the time taken to search the path using the modified bidirectional search was 1.17 ms. The same path was searched using modified DFS in 5.74 ms. Figures \ref{fig4a} and \ref{fig4b} shows two paths searched between source-target pairs 'oe23' and 'of42' (and 'oe33' and 'of01'). Although the modified bidirectional search algorithm is faster than the modified DFS while searching the shortest path, it fails while accessing parallel nodes of the same unit as input and output, as shown in Figure \ref{fig4c} (marked in red oval). An intersection is found at the node 'oe0', which is the corner node of MZI. Paths are created from this node, we can see both paths in the forward and backward direction from the source, and target nodes are physical, but the overall path is nonphysical. This problem is not encountered in modified DFS, shown in Figure \ref{fig4d}. Mean exectuion time for DFS shortest path search algorithm for 5-cell and 11-cell networks were calculated to be 7.16 ms and 88 ms, respectively. A 7-cell architecture similar to ones reported is \cite{ref20} was also analyzed and the mean execution time for DFS shortest search was 7.52 ms.

\subsection{Single Source-Target All Paths}
\begin{figure*}[!t]
\centering
\subfloat[]{\includegraphics[width=2.5in]{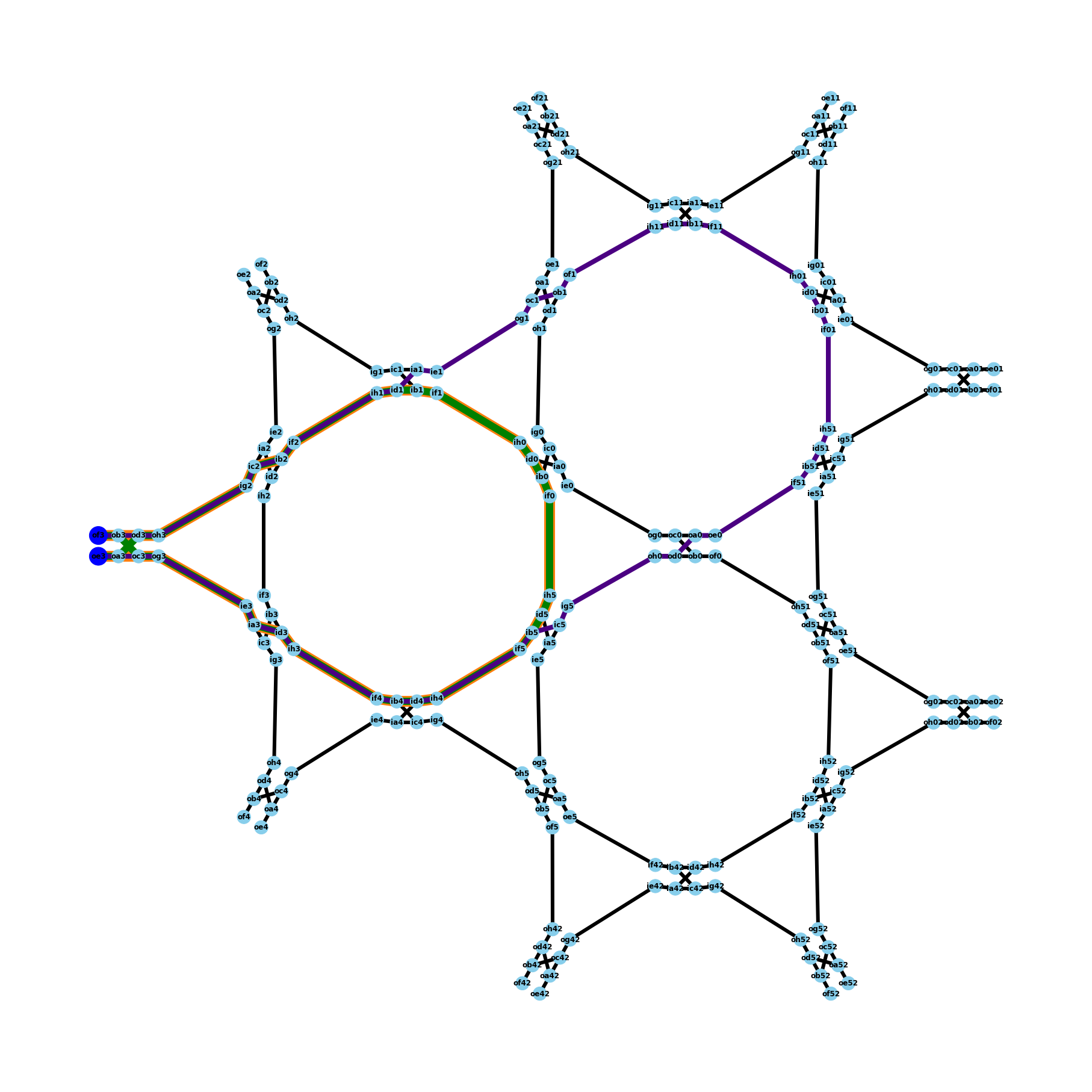}%
\label{fig5a}}
\hfil
\subfloat[]{\includegraphics[width=2.5in]{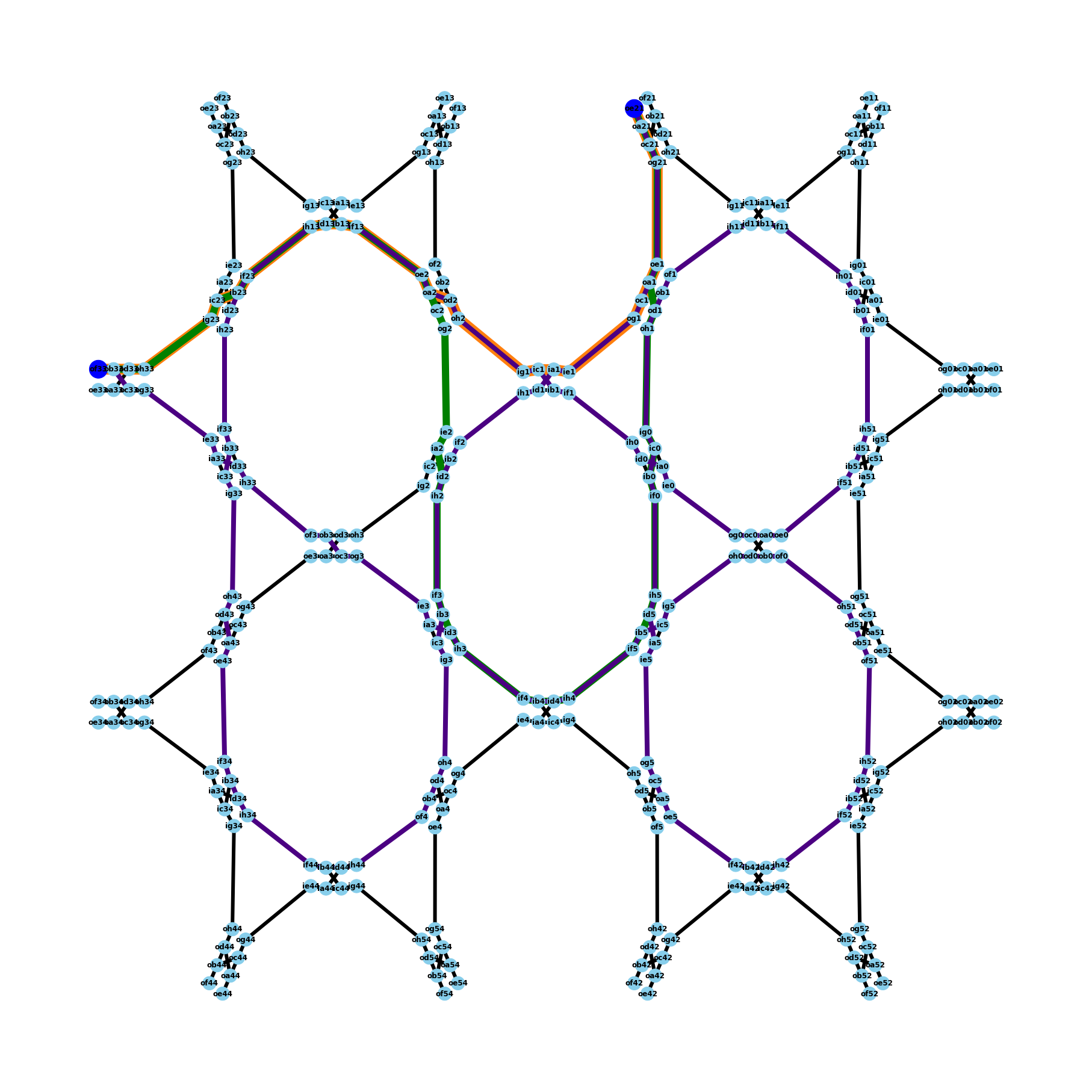}%
\label{fig5b}}
\caption{All paths between source and target using DFS (a) case I  'of3'-'oe3' with weight 8 (orange), 8 (green) and 12 (violet) and (b) case II 'of33'-'oe21' with weight 7 (orange), 11 (green) and 33 (violet).}
\label{fig5}
\end{figure*}

DFS search is also employed for finding all the existing physical paths between a pair of source and target. Case I in Figure \ref{fig5a} shows the path existing between the two parallel nodes of the same unit, 'of3'-'oe3' in a 3 unit cells network. A total of 20 paths were found, time taken to find these paths was 4 ms figure shows 3 of these paths with weights of 8 (red), 8 (green) and 12 (violet). In case II, shown in Figure \ref{fig5b}, paths were found between 'of33' and 'oe21'. A total of 484 paths with weights ranging from 7 to 33 were found, and the time taken to search these paths was 218.3 ms. The Figure shows 3 of these paths with weights of 7 (orange), 11 (green) and 33 (violet). Mean execution time to search paths over different source target pairs in a 5-cell network was 220 ms.  

\subsection{Multiple Source-Target Path}
\begin{figure*}[]
\centering
\subfloat[]{\includegraphics[width=2.7in]{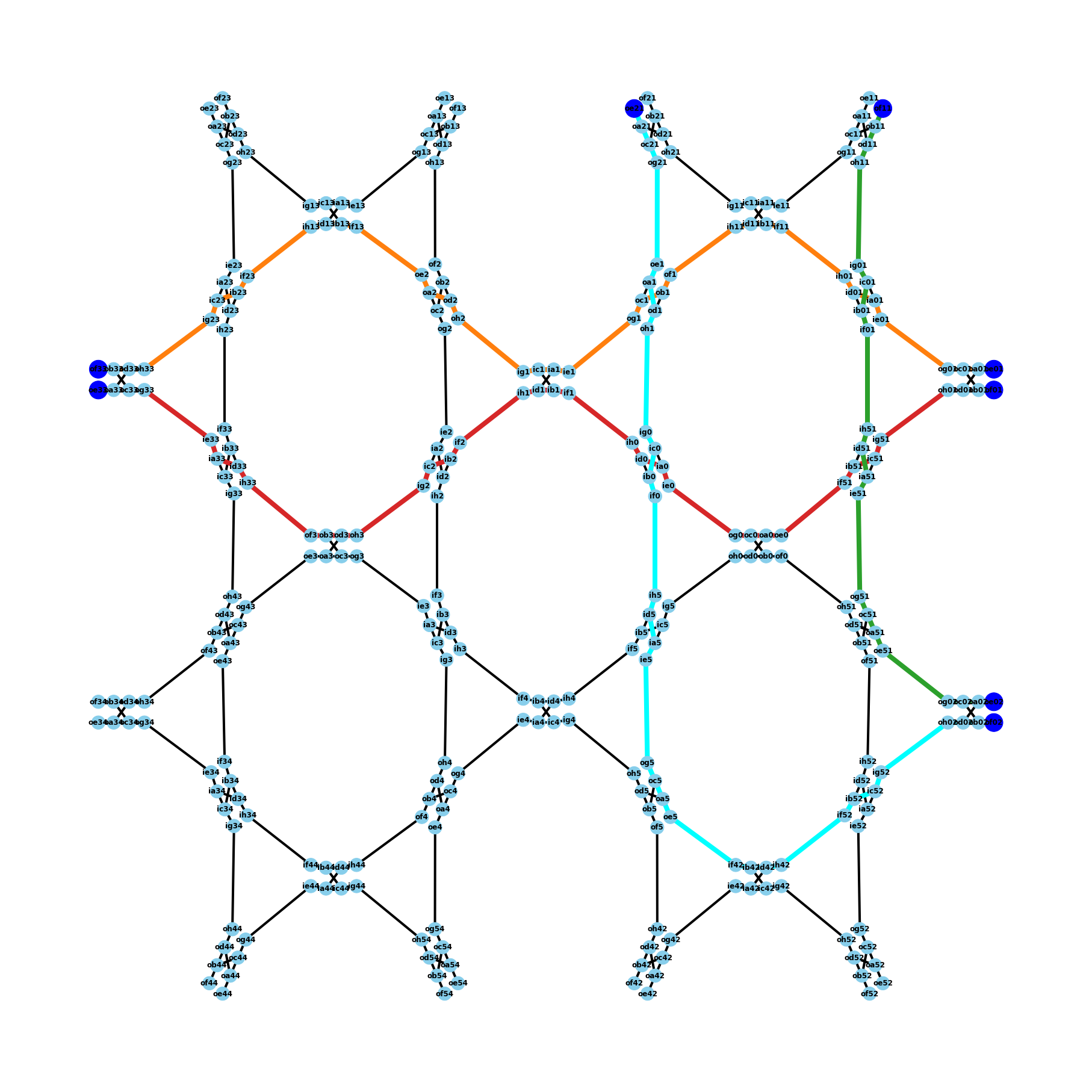}%
\label{fig6a}}
\hfil
\subfloat[]{\includegraphics[width=2.7in]{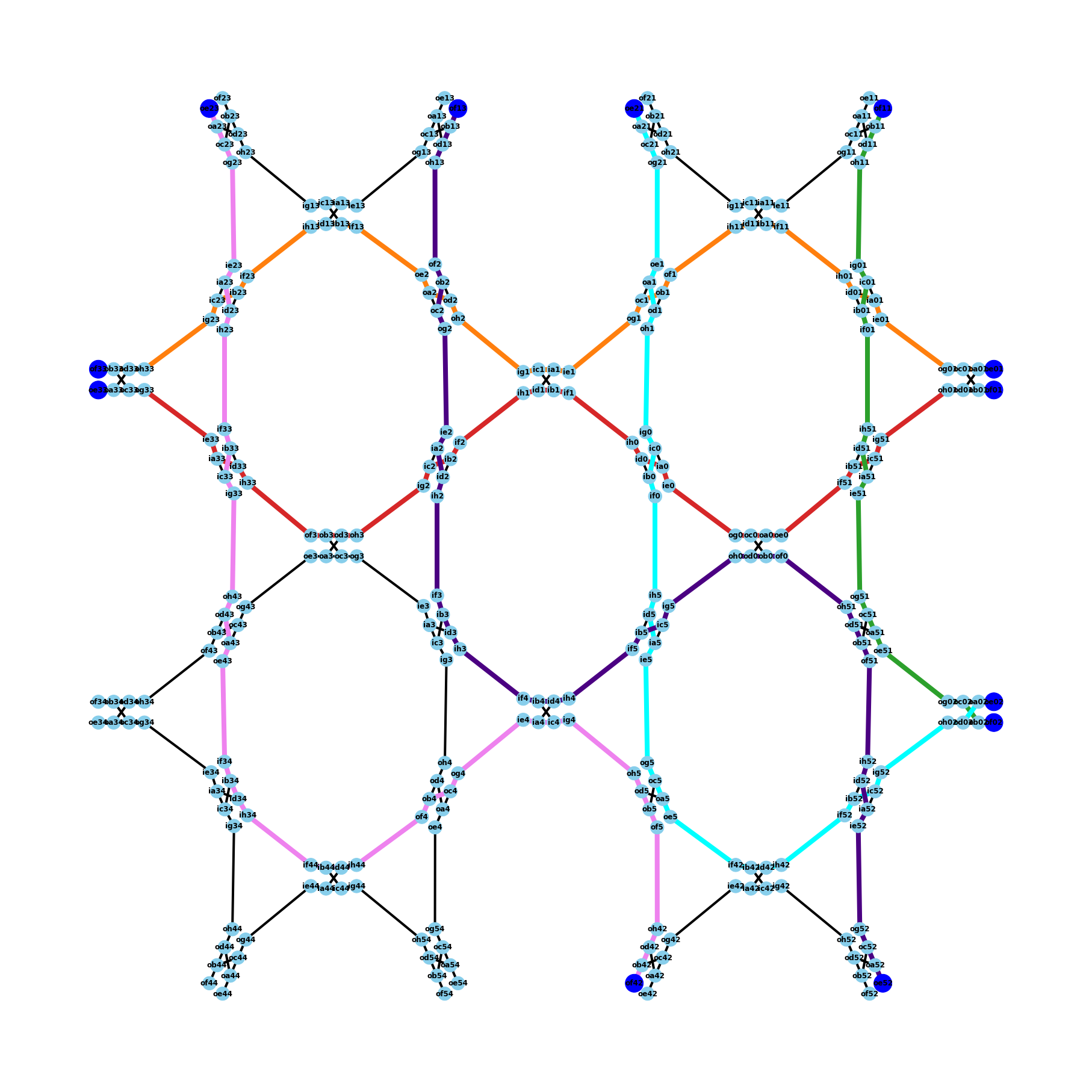}%
\label{fig6b}}
\hfil
\subfloat[]{\includegraphics[width=2.7in]{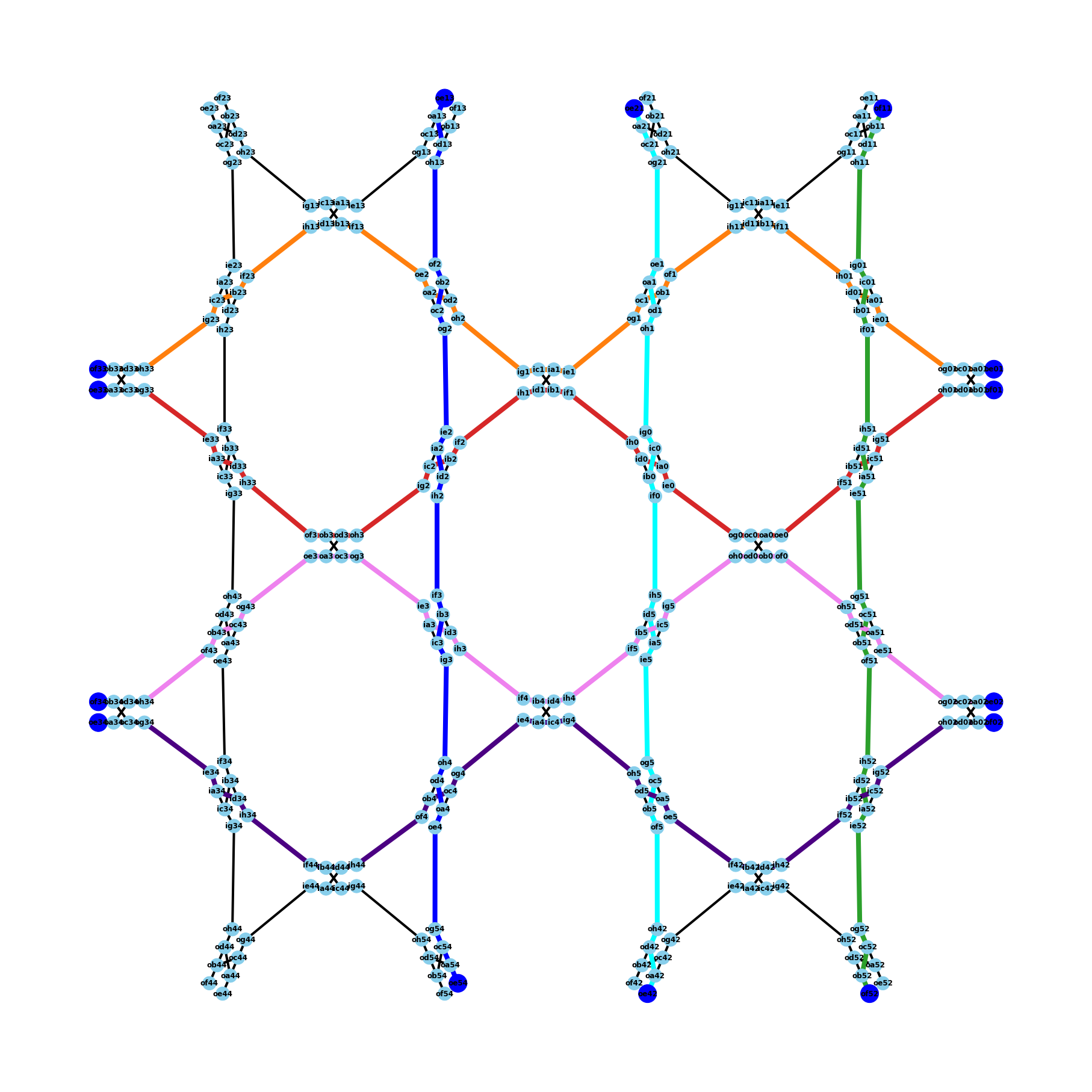}%
\label{fig6c}}
\hfil
\subfloat[]{\includegraphics[width=2.7in]{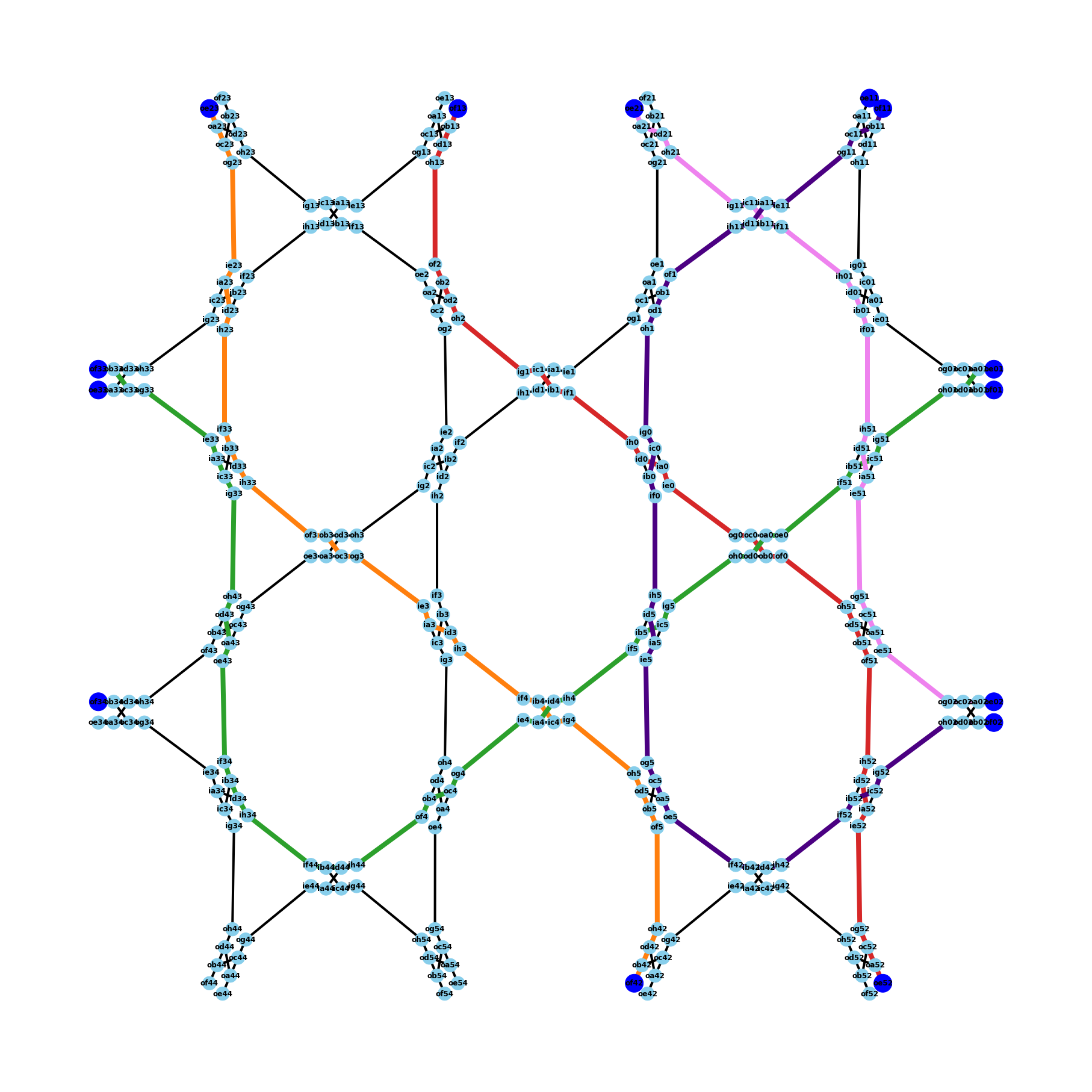}%
\label{fig6d}}
\hfil
\subfloat[]{\includegraphics[width=2.7in]{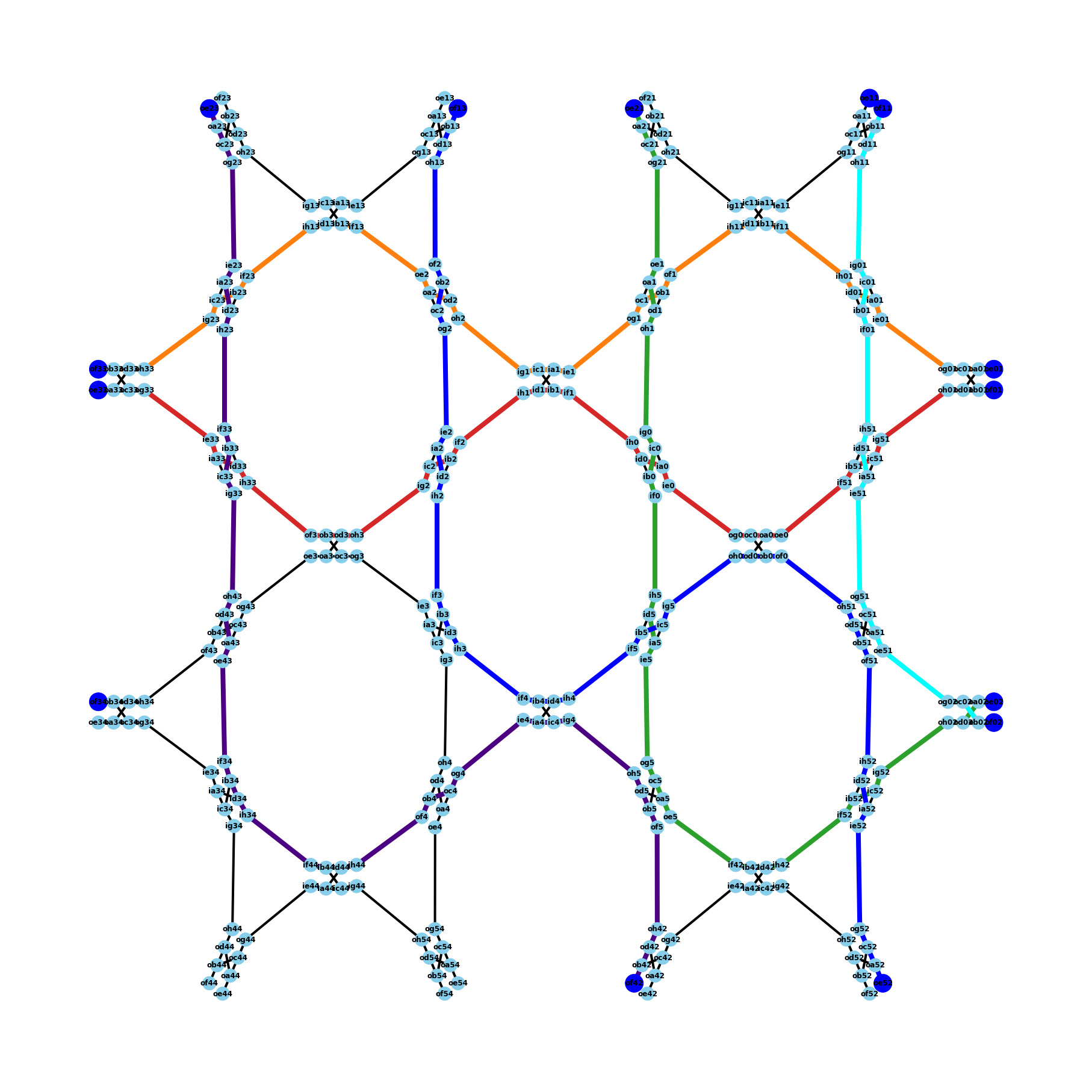}%
\label{fig6e}}
\hfil
\subfloat[]{\includegraphics[width=2.7in]{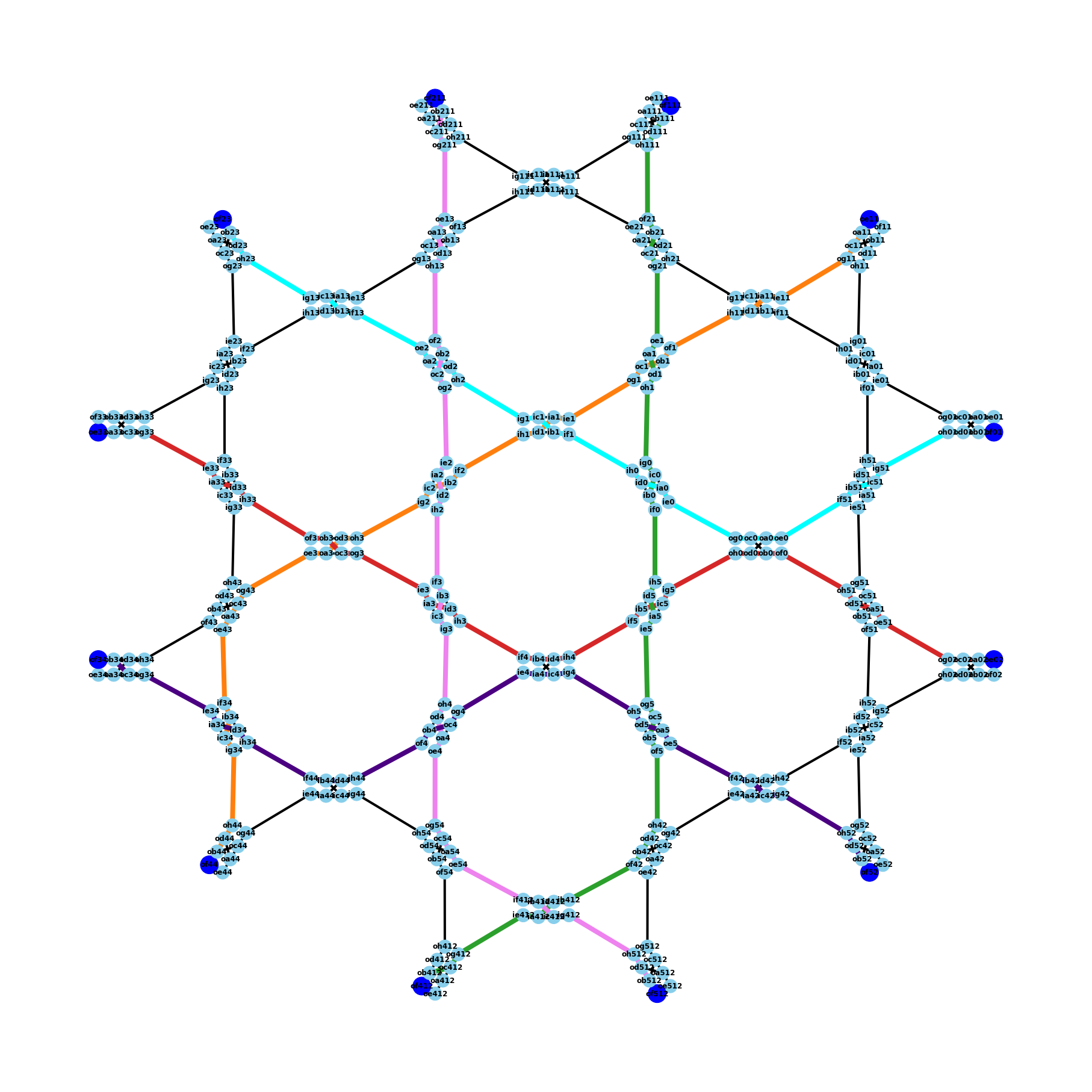}%
\label{fig6f}}
\caption{Multiple source-target paths (a) case I with four paths, (b) case II with six paths, (c) case III with seven paths, (d) and (e) case with different order of source-target pairs, and (f) six paths in a 7 cell architecture similar to \cite{ref20}.}
\label{fig6}
\end{figure*}

Due to higher accuracy, modified DFS is employed to find the shortest path between multiple source-target pairs. To prevent traversing of the same unit twice while path search and to maintain bar cross condition, the visited list is updated with the list of nodes engaged in the previous paths. Three different cases were taken for finding 4,6, and 7 paths between different pairs of source-target nodes in the network. In the first case, paths were found between 'of33'-'oe01', 'oe33'-'of01', 'of11'-'oe02' and 'oe21'-'of02', as shown in Figure \ref{fig6a}. The time taken to search these paths was 16.7 ms. Here we can observe that in comparison to the path found between the nodes 'oe33'-'of01' in the single source-target shortest path case, the path found in this case is different. Since a path exists between the nodes 'of33'-'oe01', an appropriate path is found to prevent violation of the bar and cross condition of the individual unit. In case II, paths were found between the nodes 'of33'-'oe01', 'oe33'-'of01', 'of11'-'of02', 'oe21'-'oe02', 'oe23'-'of42' and 'of13'-'oe52', as shown in Figure \ref{fig6b}. Here again, while comparing the paths obtained for the nodes 'oe23'-'of42' in the single source-target shortest path and multi source-target shortest path, we can observe that instead of getting the shortest path as in the prior case, a path maintaining the bar and cross condition for individual units was found. The time taken to search 6 paths was 17.8 ms. Figure \ref{fig6c} shows case III, where paths are found between the nodes 'of33'-'oe01', 'oe33'-'of01', 'of11'-'of52', 'oe21'-'oe42', 'of34'-'oe02', 'oe34'-'of02' and 'oe13'-'oe54'. Here time taken to search 7 paths was 18.4 ms. To further evaluate the accuracy of the algorithm, another case of searching paths between 7 source-target pairs was taken. This case was run two times with different orders of source-target pairs. First order was 'oe23'-'of42', 'of13'-'oe52', 'of33'-'oe01', 'oe33'-'of01', 'oe21'-'oe02', 'of11'-'oe52' and 'of34'-'oe11', and the second order was 'of33'-'oe01', 'oe33'-'of01', 'oe21'-'oe02', 'of11'-'of02', 'of34'-'oe11', 'oe23'-'of42' and 'of13'-'oe52'. Path obtained for the two cases are shown in Figures \ref{fig6d} and \ref{fig6e}, respectively. Since searching paths between multiple source-target pairs is a sequential process where paths are searched in the given order of source-target pairs, we can observe that for the first order, 5 paths were found out of 7; however, for second order, 6 paths were found out of 7. In all the paths found here, no nonphysical path or violation of the bar cross condition was observed, the algorithm returned empty if no physical path existed, which proves the accuracy of the algorithm. A 7 cell structure similar to one reported in \cite{ref19}, \cite{ref20} have also been analyzed in \ref{fig6f} and execution time for searching 6 paths between 'of44'-'oe11', 'oe33'-'oe02', 'of111'-'of412', 'of23'-'of01', 'of211'-'of512' and 'of34'-'of52' was 18.5 ms. For the same network 8 paths were searched in the exectuion time of 19.1 ms.

\subsection{Cycle Search From a given Parent Node}
The cycle search algorithm is employed to search all the cycles from a given parent node. The parent node is selected as the starting node of any MZI. Two cases, one for a smaller network of three cells and another for a larger network, are considered. For the case I shown in Figure \ref{fig7a}, all possible cycles with 'ih2' as the parent node were searched, and ten cycles with weights ranging from 6 to 18 were found. The Figure shows two of these cycles with weights 6 (orange) and 10 (green). The network consists of 24 MZIs, and the time taken to search all the cycles was 7.56 ms. It is worth noting that while searching total of 20 cycles are found, accounting for search in both the directions from the parent and then cycles with the same nodes are removed. In case II, all the cycles were searched along the node 'ih33' in the larger network, and 96 cycles with weights ranging from 6 to 30 are found. Figure \ref{fig7b} shows the cycles of weight 6 (orange) and 10 (green), and Figure \ref{fig7c} shows the cycle of weight 30. From the Figure, it is observed that no nonphysical path is traversed in any of the cycles, which proves the accuracy of the cycle search algorithm. Mean exectution time ofcycle search algorithm for various parent nodes was calculated to be 163.9 ms.

\subsection{N\texttimes N Photonic Switching Network}
\begin{figure*}[]
    \centering
    \subfloat[]{\includegraphics[width=2.7in]{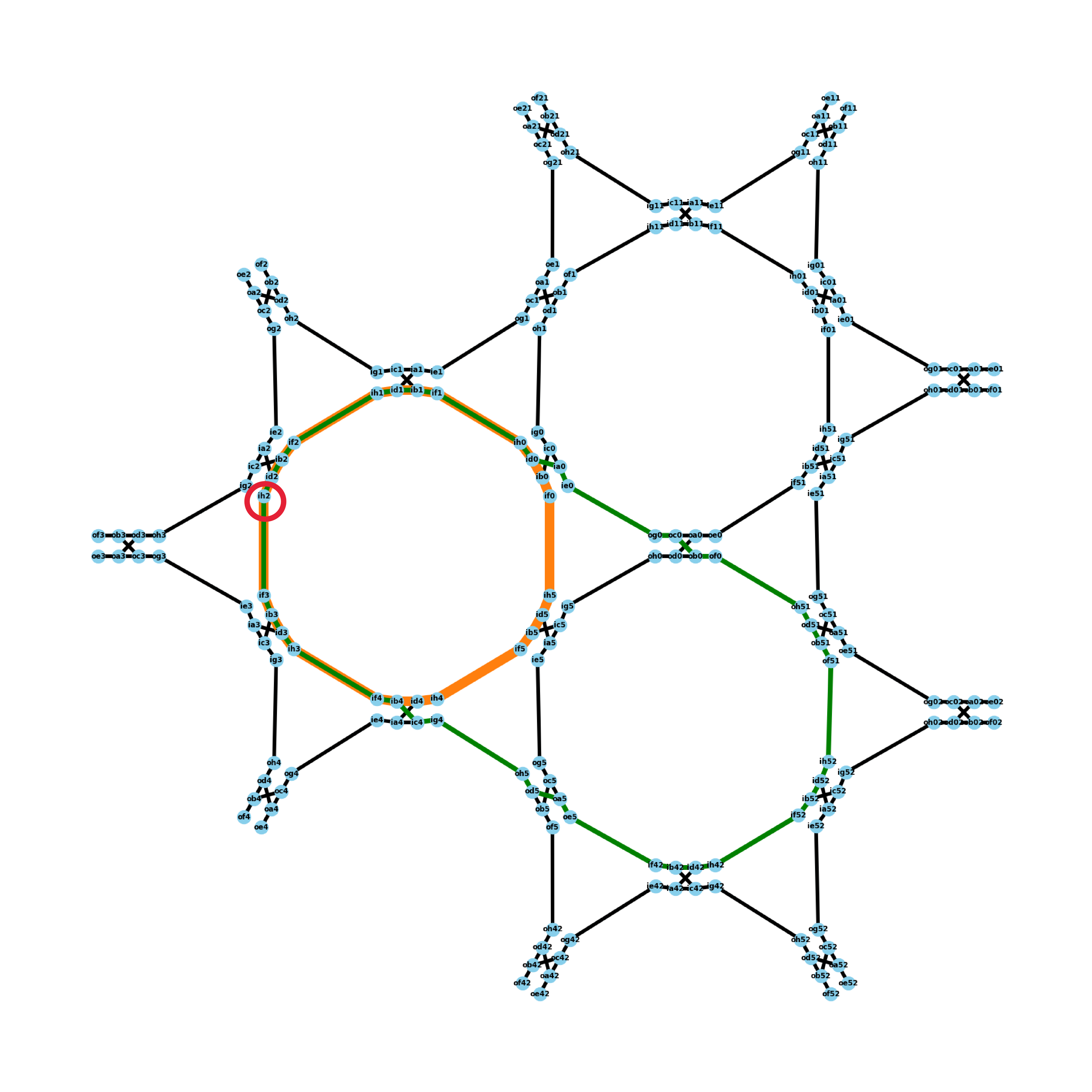}%
    \label{fig7a}}
    \hfil
    \subfloat[]{\includegraphics[width=2.7in]{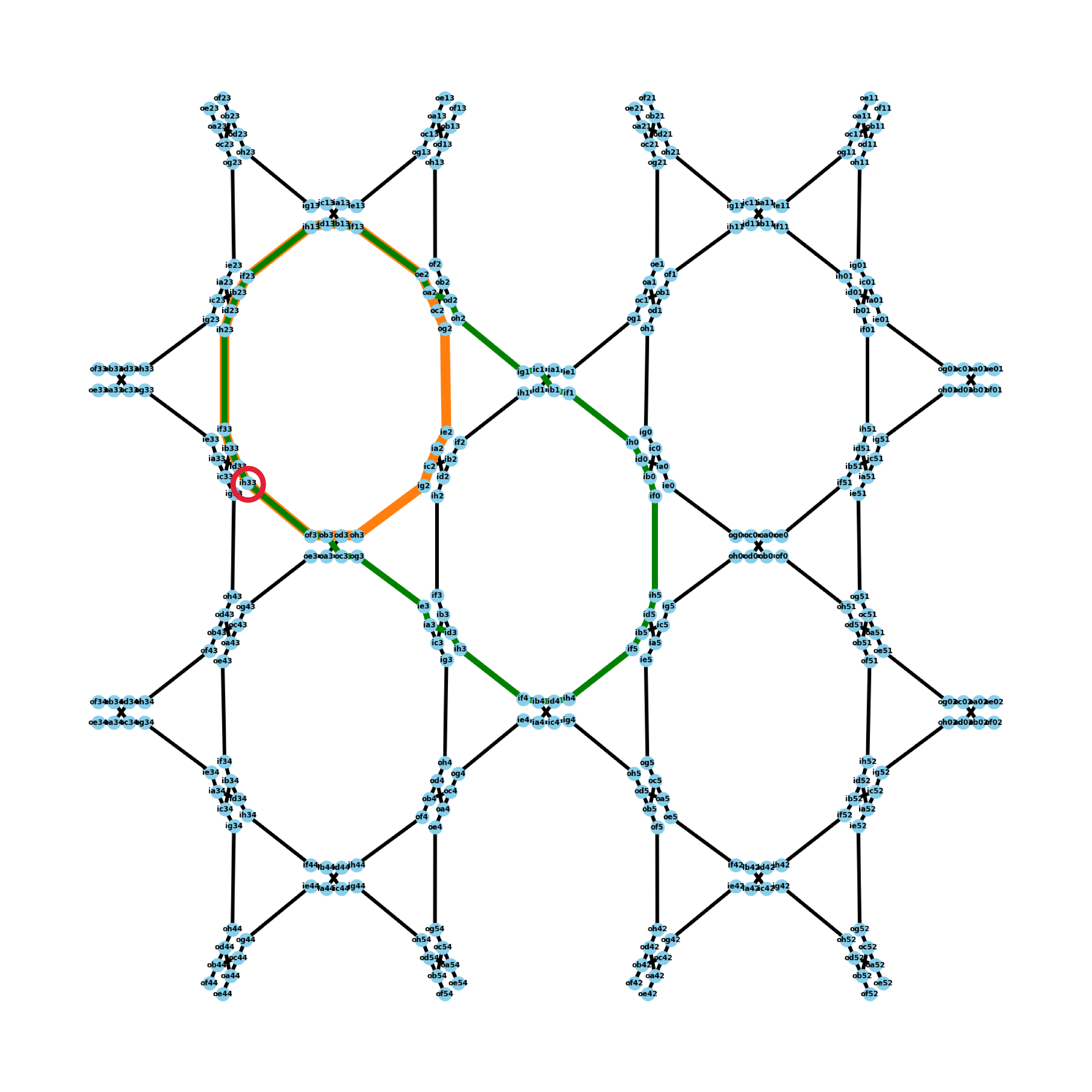}%
    \label{fig7b}}
    \hfil
    \subfloat[]{\includegraphics[width=2.7in]{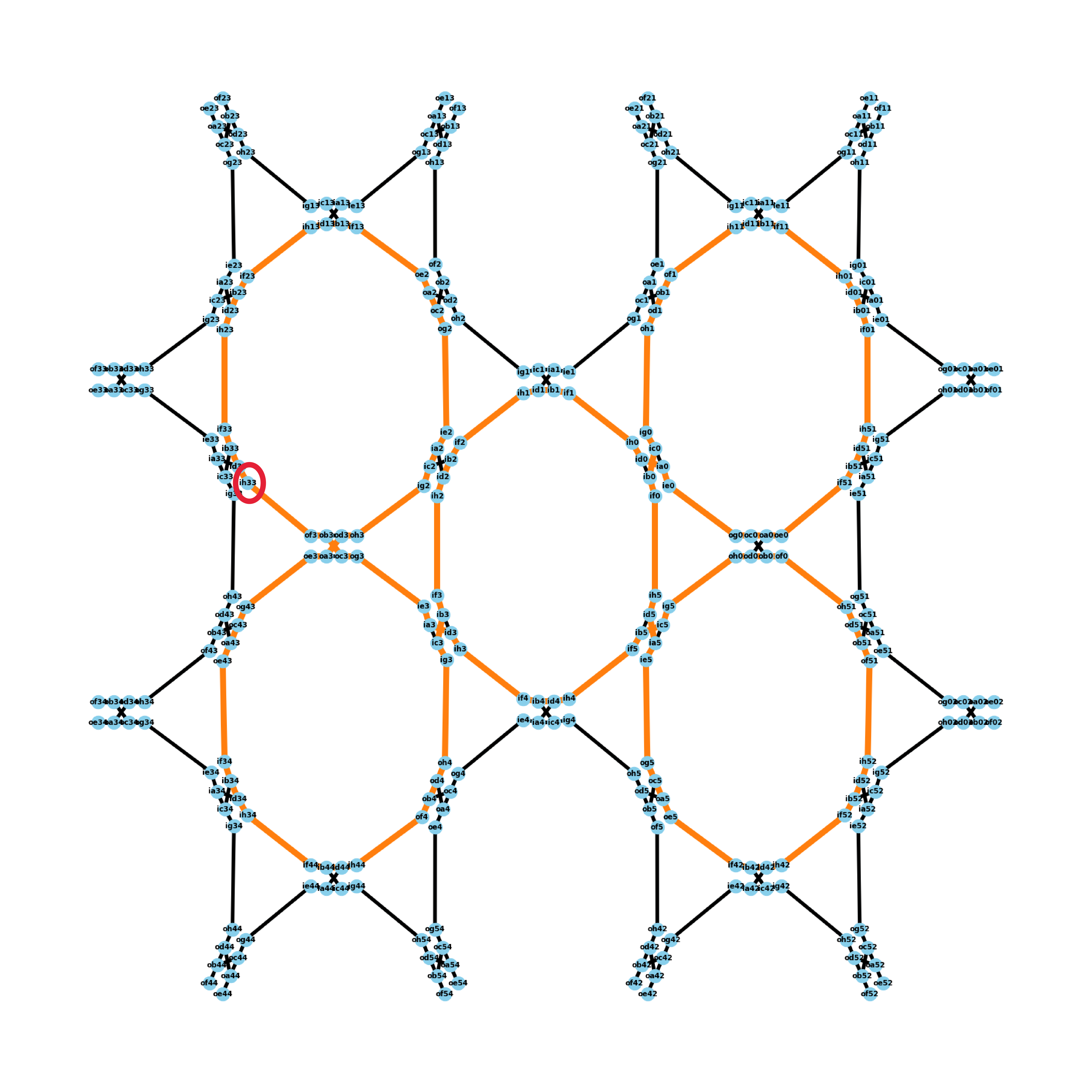}%
    \label{fig7c}}
    \hfil
    \subfloat[]{\includegraphics[width=2.7in]{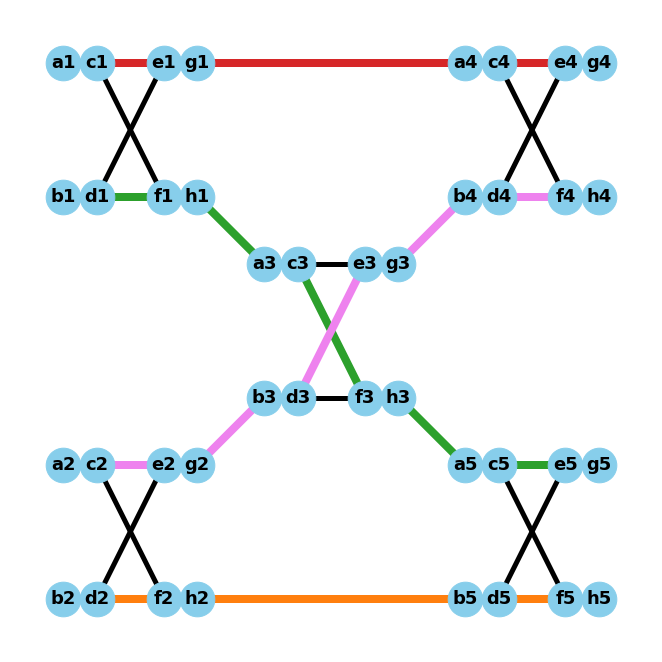}%
    \label{fig7d}}
    \hfil
    \caption{Cycle search algorithm (a) case I cycles from parent node 'ih2' with weight 6 (orange) and 10 (green), (b) case II cycles from parent node 'ih33' with weight 6 (orange) and 10 (green) (c) case II cycle of weight 30 from the parent node 'ih33'(orange), starting nodes represented in the red circles and (d) possible combination of a $4 \times 4$ switching network found using Modified DFS algorithm.}
    \label{fig7}
    \end{figure*}

N\texttimes N photonic switching networks are commonly employed for fast switching at data centres. DFS algorithm can also be used to find all the possible combinations of paths available in these networks. Two methods can be applied to find these paths, one is by listing all the paths between the inputs and outputs and saving them to the hash list, and another is dynamically searching the path given a few required input and output combinations. In both cases, the requirement is to ensure that each unit's bar and cross condition remain intact. Since the DFS algorithm proves to be highly accurate, the condition always remains intact. Figure \ref{fig7d} shows one such possible combination of input and outputs found in a 4$\times$4 switching network. 

\subsection{Eliminating the Malfunctioning Units}
Cases of malfunctioning units are very common in a large network, so it becomes necessary to eliminate those units from the path search. Malfunctioning may occur in terms of power consumption, non-functional unit, or insertion loss. Weights in each unit can be modified to correspond to any of these terms or a combination of these terms. Since the cost of traversing a physical path in an individual unit is 1, thus complexity in assigning weights in terms of either malfunctioning term becomes easy. Based on these weighted units, if any individual unit malfunctions, it can be provided with a higher weight, and then the path may be optimized for the required purpose. Another way of eliminating these malfunctioning units is to remove the nodes of the malfunctioning unit from the graph, thus not allowing the search algorithms to encounter the nodes in a malfunctioning unit. Both these ways are accurate but demand changes to be made in the graph either in terms of weights or in terms of removal of nodes which is a time-consuming process. Another way to counter the traversing through a malfunctioning unit is to simply add the nodes corresponding to the malfunctioning unit to the visited list in the search algorithms, thus preventing the algorithms from traversing through these units. This method is much more dynamic and faster than the first two methods but limits the network from accessing these units entirely. 
\begin{table*}[h]
	\begin{center}
	\caption{Performance comparison table for various algorithms in terms of complexity, maximum units traversed (longest path), the number of paths found, and mean/ total execution time taken in the search.}
	\label{tab1}
	\begin{tabular}{|c|c|c|c|c|c|}
	\hline
	\centering
	\multirow{2}{10em}{\centering Algorithm} & \multirow{2}{5em}{Complexity} & \multirow{2}{5em}{Total MZIs} & \multirow{2}{7em}{\centering Maximum MZIs Traversed} & \multirow{2}{7em}{\centering Number of Paths Searched} & \multirow{2}{5em}{Time Taken}\\ 
	& & & & & \\
	\hline
	\hline
	\multirow{4}{4em}{\centering Modified Dijkstra} & \multirow{4}{5em}{\centering O(V\textsuperscript{2})} & 30 & 9 & 1 & 56 ms\\
	& & 81 & 14 & 1 & 703 ms \\
	& & 42 & 9 & 1 & 31 ms \\
	& & 42 & 9 & 6 & 78.9 s \\
	\hline
	\multirow{2}{10em}{\centering Modified Bidirectional Search} & \multirow{2}{5em}{\centering O(2\textsuperscript{d/2})} & 36 & 9 & 1 & 1.13 ms\\
	& & 36 & 10 & 7 & 7.91 ms \\
	\hline
	\multirow{7}{4em}{\centering Modified DFS} & \multirow{7}{5em}{\centering O(V+E)} & 36 & 9 & 1 & 7.2 ms\\
	& & 36 & 11 & 7 & 18.4 ms \\
	& & 36 & 31 & 484 & 220 ms \\
	& & 42 & 9 & 1 & 7.5 ms \\
	& & 42 & 9 & 6 & 18.5 ms \\
	& & 42 & 9 & 8 & 19.1 ms \\
	& & 70 & 17 & 1 & 88 ms \\
	& & 70 & 17 & 6 & 320 ms \\
	\hline
	Cycle Search & O(V+E) & 36 & 30 & 96 & 163.9 ms \\
	\hline
	\end{tabular}
	\end{center}
	\end{table*}
\section{Discussion and Conclusion}
The algorithms discussed here provide fast and dynamically controlled routing routines for arbitrary photonic networks allowing the search for the shortest path, all paths and cycles in a given network. The proposed algorithm can be further extended to search a paths of a desired length by varying the return condition in proposed shortest path algorithm. Using these algorithms, various functionalities such as arbitrary delay lines, beam forming networks, linearized modulation units, resonant architectures, higher order filters etc., can be programmed in general-purpose photonic processors. Complexity, maximum units (MZIs) traversed in the path/ paths searched and mean execution time (total execution time in case of multipath search) for searching the paths through the different networks for various algorithms is listed in Table \ref{tab1}. The performance of proposed algorithms in terms of the mean execution time required to find paths/ cycles has been calculated using the Google Collab servers. Performance can be further enhanced using more robust processors. We also analyzed time consumption using an intel core i7 7\textsuperscript{th} Gen laptop Processor (12 Gb RAM, clock frequency 2.8 GHz) and observed that the time consumption of algorithms was reduced to half compared to Collab servers. Observations in Table \ref{tab1} show that a single path search is executed in 7.2 ms and a multipath search is executed in 18 ms (which is 2.5 times the speed of a single path search), which indicates that the algorithm speed scales up as paths are searched in the multipath search. Similar observations are made for an extensive network where 6 paths were searched in 320 ms. Another way to enhance the speed of these search algorithms is to create a hash list consisting of all the paths between each input and output using proposed DFS all path search algorithm. This will reduce the complexity of the graph search to O(E). However, each time a path is engaged, it needs to be updated in the hash list by creating a copy without engaged nodes, which might affect the performance. Methods to eliminate or counter the malfunctioning units have also been discussed, where the user can either eliminate the malfunctioning unit from the search by listing the nodes of such units in the visited list or by modifying the weight of the malfunctioning unit allowing the user to choose the path consisting of the malfunctioning unit if required. 

{\appendix[Algorithms]
A graph is defined using network-X library in python. Singular unit cells are defined through a graph name 'GX-1' where 'X' is the cell number and thus network created using 'X' cells is defined as 'NX', for example 'N2' means a graph of 3 cells with individual unit cells defined as 'G0','G1' and 'G2'.
\section*{Modified Depth First Search}
Modified DFS in all its version takes four mandatory inputs: (i) Graph, (ii) Source Node, (iii) Target Node and (iv) Visited List. In case of cycle search the source and target node gets the same entry i.e., parent node.

\begin{algorithm}[H]
\caption{Modified DFS All Paths}\label{alg:alg1}
\begin{algorithmic}
\STATE \textbf{RecursiveFun}:(curr, target, visited, all\textunderscore path,path, temp\textunderscore weight)
\STATE \hspace{0.5cm} add current to visited and path
\STATE \hspace{0.5cm} add weight to temp\textunderscore weight
\STATE \hspace{0.5cm}  If(curr $=$ target):
\STATE \hspace{0.75cm} add path to all\textunderscore paths
\STATE \hspace{0.75cm} add temp\textunderscore weight to total\textunderscore weight
\STATE  \hspace{0.5cm} Else
\STATE \hspace{0.75cm} for next node in graph[curr]:
\STATE \hspace{1cm} if temp\textunderscore Weight$>$threshold: break
\STATE \hspace{1cm} if curr not in visited: recursively call self
\STATE \hspace{0.5cm} remove u from visited and temp\textunderscore weight$=$0
\STATE \textbf{return} all\textunderscore paths and total\textunderscore weight
\STATE {{\textsc{\textbf{fun}}}: (graph, source, target, visited)}
\STATE \hspace{0.5cm} \textbf{initialize:} path, temp\textunderscore weight, all\textunderscore paths, total\textunderscore weight
\STATE \hspace{0.5 cm} all\textunderscore paths, total\textunderscore weight$=$call(RecursiveFun)
\STATE \hspace{0.5 cm} Sort all\textunderscore paths and total\textunderscore weight
\end{algorithmic}
\label{alg1}
\end{algorithm}

\begin{algorithm}[H]
\caption{Modified DFS Shortest Path}\label{alg:alg2}
\begin{algorithmic}
\STATE \textbf{RecursiveFun}:(curr, target, visited, path, currpath, temp\textunderscore weight)
\STATE \hspace{0.5cm} add current to visited and currpath
\STATE \hspace{0.5cm} add weight to temp\textunderscore weight
\STATE \hspace{0.5cm}  If(curr $=$ target):
\STATE \hspace{0.75cm} make currpath $=$ paths
\STATE \hspace{0.75cm} make temp\textunderscore weight $=$ total\textunderscore weight
\STATE  \hspace{0.5cm} Else
\STATE \hspace{0.75cm} for next node in graph[curr]:
\STATE \hspace{1cm} if temp\textunderscore Weight$>$threshold: break
\STATE \hspace{1cm} if weight(currpath)$>$weight(path): break
\STATE \hspace{1cm} if curr not in visited: recursively call self
\STATE \hspace{0.5cm} remove u from visited and temp\textunderscore weight$=$0
\STATE \textbf{return} path and total\textunderscore weight
\STATE {{\textsc{\textbf{fun}}}: (graph, source, target, visited)}
\STATE \hspace{0.5cm} \textbf{initialize:} path, temp\textunderscore weight, path, total\textunderscore weight
\STATE \hspace{0.5 cm} path, total\textunderscore weight$=$call(RecursiveFun)
\end{algorithmic}
\label{alg2}
\end{algorithm}

\begin{algorithm}[H]
\caption{Modified DFS Cycle Search}\label{alg:alg3}
\begin{algorithmic}
\STATE \textbf{CycFun}:(curr, par, visited, curr\textunderscore cycle, all\textunderscore cycles, temp\textunderscore weight)
\STATE \hspace{0.5cm} add current to visited and cycle
\STATE \hspace{0.5cm} add weight to temp\textunderscore weight
\STATE \hspace{0.5cm}  for next node in graph[curr]
\STATE \hspace{0.75cm} if temp\textunderscore Weight$>$threshold: break
\STATE \hspace{0.75cm} if curr not in visited: recursively call self
\STATE \hspace{0.75cm} elseif curr in visited and curr$=$ par:
\STATE \hspace{1cm} add cycle to all\textunderscore cycles
\STATE \hspace{1cm} add temp\textunderscore weight to total\textunderscore weight
\STATE \hspace{0.5cm} remove u from visited and temp\textunderscore weight$=$0
\STATE return all\textunderscore cycles and total\textunderscore weight
\STATE {{\textsc{\textbf{fun}}}: (graph, par, par, visited)}
\STATE \hspace{0.5cm} \textbf{initialize:} cycle, temp\textunderscore weight,all\textunderscore cycles, total\textunderscore weight
\STATE \hspace{0.5 cm} all\textunderscore cycles, total\textunderscore weight$=$call(CycFun)
\STATE \hspace{0.5 cm} Sort all\textunderscore cycles and total\textunderscore weight
\end{algorithmic}
\label{alg3}
\end{algorithm}

}

\section*{Acknowledgments}
We thank the Government of India for funding this research under Prime Minister's Research Fellowship Scheme. We would also like to thank Ms Pragya Mishra from Applied Photonics Lab, Department of Electrical Communication Engineering, Indian Institute of Science, Bangalore, for thoughtful discussions and for reviewing the manuscript.

\vfill

\end{document}